\documentclass[smallcondensed]{svjour3}     
\smartqed  
\usepackage{graphicx}
\usepackage{hyperref}
\usepackage{bbold}
\usepackage{mathtools}
\usepackage{amsmath}
\usepackage{amsfonts}
\usepackage{tikz}
\usepackage[utf8]{inputenc}
\usetikzlibrary{arrows, decorations.pathmorphing}
\newcommand{\cO}{{\cal O}}
\newcommand{\cP}{{\cal P}}

\newcommand{\ket}[1]{|#1\rangle}
\newcommand{\bra}[1]{\langle#1|}
\newcommand{\bracket}[2]{\langle#1|#2\rangle}
\newcommand{\vev}[1]{\langle#1\rangle}
\newcommand{\h}{\mathfrak{q}}
\renewcommand{\S}{\textsc{s}}
\newcommand{\I}{\textsc{i}}
\newcommand{\R}{\textsc{r}}
\newcommand{\E}{\textsc{e}}
\newcommand{\X}{\textsc{x}}
\newcommand{\Y}{\textsc{y}}
\newcommand{\A}{\textsc{a}}

\newcommand{\hSI}{\mathfrak{q}_{\S \to \I}}
\newcommand{\hIS}{\mathfrak{q}_{\I \to \S}}
\newcommand{\hIR}{\mathfrak{q}_{\I \to \R}}
\newcommand{\hXY}{\mathfrak{q}_{\X \to \Y}}
\newcommand{\StoI}{\S \to \I}
\newcommand{\SIS}{\S\I\S}
\newcommand{\SIR}{\S\I\R}
\newcommand{\Q}{\mathbb{Q}}

\begin{document}

\title{Exact epidemic models from a tensor product formulation
}


\author{Wout Merbis}


\institute{W. Merbis \at
	          Dutch Institute for Emergent Phenomena (DIEP), Institute for Theoretical Physics, University of Amsterdam, 1090 GL Amsterdam, The Netherlands \\
			  Physique Th\'eorique et Math\'ematique, 
              Universit\'e Libre de Bruxelles and International Solvay Institutes,
              Campus Plaine - CP 231,
              B-1050 Bruxelles, Belgium             \\
	          \email{w.merbis@uva.nl} 
}

 \date{ }

\maketitle

\begin{abstract}
A general framework for obtaining exact transition rate matrices for stochastic systems on networks is presented and applied to many well-known compartmental models of epidemiology. The state of the population is described as a vector in the tensor product space of $N$ individual probability vector spaces, whose dimension equals the number of compartments of the epidemiological model $n_c$. The transition rate matrix for the $n_c^N$-dimensional Markov chain is obtained by taking suitable linear combinations of tensor products of $n_c$-dimensional matrices. The resulting transition rate matrix is a sum over bilocal linear operators, which gives insight in the microscopic dynamics of the system. The more familiar and non-linear node-based mean-field approximations are recovered by restricting the exact models to uncorrelated (separable) states. 
We show how the exact transition rate matrix for the susceptible-infected (\S\I) model can be used to find analytic solutions for $\S\I$ outbreaks on trees and the cycle graph for finite $N$. 
\end{abstract}


\section{Introduction}
\label{intro}

At all length scales in Nature, collective behavior can emerge from the interactions of many individual constituents. These interactions can be characterized by (relatively) simple rules and still give rise to complex collective behavior. At short length scales, the rules of quantum mechanics are responsible for the manifestation of exotic phases of matter such as superconductivity or Bose-Einstein condensates. At longer length scales, the activity of neurons interacting through complex connectivity networks is thought by many to determine the emergence of intelligence and consciousness. Individual intelligent agents in turn interact to forge great collaborative works, but also risk the chance of creating epidemic outbreaks, such as the Covid-19 pandemic the world is facing today. 

Significant progress in the field of complex systems has shown that complex connectivity networks underly broad classes of emergent phenomena \cite{albert2002statistical,boccaletti2006complex}. On the one hand, this discovery led to a tremendous experimental effort aiming at uncovering the hidden network structures of real-world complex systems, while, on the other hand, it triggered a series of numerical developments with the purpose of simulating complex systems on these networks. Within the context of these systems, emergent properties are often studied using a simplified description in terms of probabilistic (stochastic) models. This is well-exemplified in the modeling of epidemic spreading on networks, where individuals are treated as a set of random variables taking values in a specified set of states (for instance, ‘infected’ or ‘susceptible’). Theoretical models based on Markov chains govern transitions between these states \cite{PastorSatorras2015,kiss2017mathematics,newman2010networks}. Despite the large body of work devoted to numerical simulations and mean-field approximations, a complete analytical understanding of these `many-body stochastic systems’ is still lacking. In this paper, we make progress in this direction by taking inspiration from many-body quantum systems and formulate the exact Markov chain for many models of epidemic spreading on network in terms of bilocal operators acting on the tensor product space of probability vectors.

The mathematical description of the epidemic spreading process due to Kermack and McKendrick \cite{kermack1927} laid the foundation for the development of \textit{compartmental models} of epidemiology, which has since grown into an important branch of population dynamics \cite{Anderson1992,keeling2011modeling,Brauer2012}. The population of interest is divided into $n_c$ different compartments, such as susceptible ($\S$), infected ($\I$) or recovered ($\R$). These represent the possible stages an individual goes through when contracting an infectious disease. Any individual belongs to one compartment and a set of interaction rules is introduced which describe the possible transitions between compartments as a stochastic process. Popular models are the $\SIR$ model \cite{kermack1927} and the $\SIS$ model \cite{kermack1932contributions,WEISS1971261}, both containing a $\StoI$ transmission transition to model the spread of the infection, and a recovery process which assumes individuals acquire immunity for $\SIR$, but not for $\SIS$. Both models show threshold behavior in the sense that there is a critical value for the ratio of transmission and recovery rates above which the infection spreads to a non-zero fraction of the total population. This threshold is called the \textit{epidemic threshold}. Many generalizations of the $\SIR$ and $\SIS$ models exist and contain additional compartments and transitions to further refine the model according to the specifics of the pathogen \cite{Anderson1992,keeling2011modeling,Brauer2012}.

The compartmental models describe the dynamics at the level of the population and should therefore be understood as emerging from a microscopic system of individuals interacting in society. In this microscopic description, individuals are described as random variables whose expectation value gives the probability of finding this individual in any of the compartments. The deterministic rate equations of the compartmental models are recovered by taking a thermodynamic limit, which in this case means the limit of large population size $N \to \infty$. In addition, it is assumed that the population is homogeneously mixed: the interactions occur at random and each individual in the infected class has equal probability of infecting any susceptible individual. In many real-world scenarios, however, interactions between individuals form complex contact networks with large-scale heterogeneous properties \cite{albert2002statistical}. Most individuals are in direct contact with only a small part of the population, but some are able to cause super-spreading events, infecting many at the same time. These heterogeneous network properties have important effects on the epidemic threshold of the macroscopic description, see for instance \cite{pastor2001epidemic,may2001infection,pastor2002epidemic,wang2003epidemic,newman2003properties,serrano2006percolation,van2008virus}. We refer to the excellent review \cite{PastorSatorras2015} and textbooks \cite{barrat2008dynamical,newman2010networks,kiss2017mathematics} for further reading and references on this topic.

The main mathematical tools for modeling epidemic spreading on networks are numerical simulations and the use of mean-field approximations of various kinds. In the mean-field limit, the statistical independence of (subsets of) nodes is assumed. Some of the first studies of epidemic spreading on complex networks rely on a degree-based mean-field theory approximation, where all nodes of degree $k$ are considered statistically equivalent \cite{pastor2001epidemic}. Individual-based mean-field theory methods (sometimes referred to as $N$-intertwined mean-field approximation or NIMFA) assumes the statistical independence of all neighboring nodes in the network \cite{van2008virus,simon2011exact,cator2013susceptible}. This can be extended to account for non-trivial correlations between pairs of adjacent nodes, at the cost of having to track the dynamics of these pairs. In turn, this will depend on the expectation values of triples and pairs. The full set of equations can then be `closed' by assuming functional dependence of the expectation values for triples on those of pairs and singlets \cite{cator2013susceptible,kiss2015exact}. 

Mean-field approximations have been used extensively and with great success \cite{gleeson2012accuracy}, but less work has been devoted to the study of the exact Markovian system of the epidemic spreading process on complex networks. This is mainly due to the exponential scaling of the state space with population size $N$. When considering the system as a network of $N$ nodes taking $n_c$ distinct values, the exact state of the system is $n_c^N$ dimensional. For sizable populations this quickly gets out of hand; the dimensionality of a binary-compartment system with $N=300$ already exceeds the number of atoms in the observable universe. Despite this exponential complexity, some results on the exact Markov chain of epidemic models on networks have been derived. Transition rate matrices (or infinitesimal generators) for the continuous-time Markov chain have been constructed explicitly for the case of the $\SIS$ model \cite{van2008virus,cator2013susceptible} and the $\SIR$ model \cite{van2014exact}. It is also possible to obtain exact descriptions on certain symmetric networks by using graph-automorphisms to lump together suitable subsets of the exact system \cite{simon2011exact,kiss2017mathematics}. In this way, the set of $n_c^N$ coupled differential equations of the exact system can be reduced to a number of equations only polynomial in $N$.
Other exact methods focus on mapping the static properties of the $\SIR$ system to bond percolation \cite{ludwig1975final,Newman2002}.

While the enormous system size may seem like a insurmountable obstruction to making progress on the exact Markovian system, insights from many-body quantum systems indicate that the exponential complexity may be an illusion. For any many-body quantum system with bilocal interactions (involving only nearest-neighbors), it has been shown that the vast majority of the states in the composite system are not physical \cite{poulin2011quantum}. These states cannot be reached by time evolution under local interactions, at timescales polynomial in system size $N$. The subset of physical states is in fact exponentially small, making it possible to represent exact solutions of many-body quantum systems efficiently, i.e. by specifying a number of parameters only polynomial in $N$ \cite{verstraete2008matrix,orus2014practical}.
	
In this work, we take inspiration from many-body quantum systems and reformulate the exact Markovian dynamics of compartmental models of epidemiology on static networks in terms of bilocal operators acting on individual probability vectors. The state of the population is then a vector in the tensor product space of the individual probability vector spaces. Transition rate matrices $\Q$ for the exact master equation are similarly constructed from tensor products of local operators. The formulation is analogous to how local Hamiltonians are defined in many-body quantum systems, although the individual building blocks are modified appropriately to reflect the stochastic character of the problem at hand \cite{Baez_2017}. This means that instead of using Hermitian matrices to construct interaction Hamiltonians, the local operators are infinitesimally stochastic matrices, leading to a `many-body master equation' for the epidemic spreading process on networks. Due to the linearity of the exact Markovian system, the time-dependence can then be found from the matrix exponential $\exp(\Q \, t)$ of the infinitesimal generator.

While the use of tensor products to describe stochastic models is not new (see for instance \cite{felderhof1971spin,derrida1993exact}), its application to the epidemic spreading process on networks has not been explored fully in the literature so far\footnote{See, however, the appendix of \cite{sahneh2013generalized} for a closely related description of the exact $\SIS$ model.}. We make a step in this direction by illustrating how the formulation gives insight in the microscopic dynamics of the $\StoI$ transition. To this end, we study the time dependence of an $\S\I$ outbreak on finite graphs. The outbreak is initiated by placing a single infected in a completely susceptible population. By expanding the matrix exponential $\exp( \Q_{\S\I} \,t)$ in time, we can track the evolution of the outbreak as a sum over contributions from subgraphs of the network. The transition rate matrix $\Q$ imposes a set of recursive relations among the different contributions of each subgraph. These recursive relation can be solved explicitly in the case of trees and the cycle graph, leading to analytical expressions for the expectation values for individual nodes as a function of time. 

This paper is organized as follows. In section \ref{sec:tensorproduct} the tensor product formulation for stochastic systems is presented, analogously to the construction of many-body quantum Hamiltonians. For pedagogical reasons no prior knowledge of quantum mechanics is required. In section \ref{sec:Logisticgate}, the construction is applied to the $\StoI$ transition, which is an essential element of any compartmental model of epidemiology. The homogeneously mixed thermodynamic limit is shown to reproduce the Verhulst logistic differential equation \cite{verhulst1838}; a prototypical model of epidemiological spreading and population growth. Section \ref{sec:compartmentals} gives a general procedure for obtaining exact transition rate matrices for any epidemic model on networks and we work out the infinitesimal generators for $\SIS$ and $\SIR$ in detail. We briefly discuss generalizations to other compartmental models. To illustrate how the present construction can be used to gain insight into the microscopic dynamics of epidemic spreading, we turn to the pertubative expansion in time for the $\S\I$ model in section \ref{sec:perturbative}. We show explicitly how the transition rate matrices constructed in section \ref{sec:Logisticgate} can be used to find exact solutions on simple, regular graphs in section \ref{sec:exact}, focusing on the case of trees and the cycle graph. We conclude the paper with a discussion section, where we comment on the possibility of retaining a memory of past correlations, despite the Markovian nature of the system and discuss possible further applications of this mathematical technique to stochastic systems.

\section{A tensor product formulation for stochastic systems}
\label{sec:tensorproduct}

The population is modeled by a set of $N$ nodes, connected by edges taken from a graph with adjacency matrix $A^{ij}$. In realistic epidemics, the social interaction network underlying the individual contact patterns is constantly changing in time, often in response to the spreading of the infection itself. In this work, however, we will focus on the spreading process on static networks. Although less realistic, the assumption of a static interaction network increases the analytical tractability of the problem.

\subsection{Individual probability vector spaces}
\label{sec:indprop}
To each node in the network, we assign a probability vector $|\rho^i \rangle$, where $i$ labels the individual  $1 \leq i \leq N$.  This vector takes values in a set of basis states $\ket{\X}$, where $\X$ labels the compartments of the epidemiological model of interest. We denote by $C$ the set of all compartments, so $\X \in C$. For the binary compartment model with $C = \{\S , \I \}$, there are two basis states
\begin{equation}
\ket{\S} = \left( \begin{array}{c}
1 \\
0
\end{array}  \right) \,, \qquad \ket{\I} = \left( \begin{array}{c}
0 \\
1
\end{array}  \right)\,.
\end{equation}
The probability vector $|\rho^i \rangle$ is a linear combination of these deterministic basis states
\begin{equation}\label{rho2c}
\ket{\rho^i(t)} =  \rho^i_\S(t)\ket{\S} + \rho^i_\I(t) \ket{\I} = \left( \begin{array}{cc}
\rho^i_\S(t) \\
\rho^i_\I(t)
\end{array}\right)\,.
\end{equation}
In general, a model with $n_c$ compartments will give a $n_c$-dimensional probability vector. 
Since $\ket{\rho^i(t)} $ contains the probabilities of all possible states of an individual, its elements should be real, non-negative and sum to one. We will denote this sum as a contraction with the unit co-vector $\bra{1}$. In the two compartment case, this is $\bra{1} = (\,1 \; 1 \,)$ and
\begin{equation}\label{L1norm}
\bracket{1}{\rho^i(t)} = 1\,.
\end{equation}
In mathematical terms, $\ket{\rho^i(t)}$ is normalized in the $\ell^1$-norm and hence it is an element of the Lebesgue space of $\ell^1$-normalized, non-negative $n_c$-dimensional vectors, which we denote as $\cP^{n_c}$ and refer to as a probability vector space. We will refer to the elements in $\cP^{n_c}$ as \textit{stochastic states}.

When considering the individual nodes as $n_c$-dimensional probability vectors $\ket{\rho^i(t)}$, a natural operation on these objects is matrix multiplication with $n_c \times n_c$ dimensional matrices. A useful set of matrices are the projection matrices $P_\X$. There are $n_c$ of them, and they are defined by having only one non-zero entry. Specifically, the matrix $P_\X$ will be zero, except in the $(\X,\X)$ diagonal entry, which is 1, or:
\begin{equation}
P_\X = \ket{\X}\bra{\X} \,,
\end{equation}
where $\bra{\X} = \ket{\X}^T$. In the binary case, these are:
\begin{equation}
P_\S = \ket{\S} \bra{\S} = \left( \begin{array}{cc}
1 & 0 \\
0 & 0
\end{array}\right)\,, \qquad
P_\I = \ket{\I}\bra{\I} =
\left( \begin{array}{cc}
0 & 0 \\
0 & 1
\end{array}\right)\,.
\end{equation}
The projection operators are useful in computing expectation values. For instance, to extract the probability that node $i$ is infected, one can compute
\begin{equation}
	\vev{\I^i} =  \bra{1} P_\I \ket{\rho^i(t)}= \bracket{\I}{\rho^i(t)} = \rho_\I^i(t) \,.
\end{equation}
This might look like a roundabout way of saying the probability of $i$ being infected is given by its bottom component. However, in the tensor product description of the entire population, it is not always possible to simply read out the bottom component of the constituent vectors, as they might be correlated with their neighbors in the network. It is then useful to have an operator which returns the expectation value of a single node. We will show this feature in detail in the next section.

Note that the projection operators $P_\X$ are not unique. We can define the alternate projectors
\begin{equation}\label{Ptilde}
\tilde P_\S = \left( \begin{array}{cc}
0 & 0 \\
1 & 0
\end{array}\right)\,, \qquad
\tilde P_\I = 
\left( \begin{array}{cc}
0 & 1 \\
0 & 0
\end{array}\right)\,,
\end{equation}
which also satisfy the property that $\vev{\X^i} = \bra{1} \tilde P_\X \ket{\rho^i} = \rho_\X^i $. In general, for any $\Y$, the matrix $\tilde{P}_{\X} = \ket{\Y} \bra{\X}$  projects the vector $\ket{\rho}$ to its $\X$-th component under the $\ell^1$-norm.

\subsection{Transition rate matrices}
\label{sec:tranrate}
The probability vector $\ket{\rho(t)}$ evolves in time by the master equation
\begin{equation}\label{msteqn}
\partial_t \ket{\rho^i(t)} = \Q \ket{\rho^i(t)} \,.
\end{equation}
where $\Q$ is the infinitesimal generator, or the transition rate matrix\footnote{Note that many references define $\Q$ as multiplying the state vector from the right. This other convention for the infinitesimal generator can be obtained by taking the transpose of the $\Q$'s used here.}. If $\Q$ is time-independent, the solution to \eqref{msteqn} can be represented as
\begin{equation}
\ket{\rho(t)} = \exp( \Q \,t) \ket{\rho(0)}\,.
\end{equation}
In order for the total probability $\bracket{1}{\rho(t)}$ to be conserved in time, the matrix $\Q$ should satisfy the property
\begin{equation}\label{infstoch}
\bra{1} \Q = \bra{0} \,.
\end{equation}
Here $\bra{0}$ denotes the row vector with all zero entries, so all columns of the matrix $\Q$ sum to zero. In that case, $\partial_t \bracket{1}{\rho(t)}= 0$ is guaranteed by the master equation \eqref{msteqn}.
The property \eqref{infstoch} implies that the matrix exponential $S(t) = \exp(\Q\,t)$ is a (left) \emph{stochastic} matrix, with column sums equal to one:
\begin{equation}
\bra{1} S(t) = \bra{1} \,.
\end{equation}
Stochastic matrices are linear maps from stochastic states to stochastic states $S(t): \cP^{n_c} \mapsto \cP^{n_c}$. They evolve the system forward in time and their matrix elements are the conditional probabilities for each possible transition.

Transition rate matrices $\Q$ are called \emph{infinitesimally stochastic} \cite{Baez_2017} and they map stochastic states to \textit{infinitesimally stochastic states}. The latter are vectors with elements summing to zero and hence they can be added to stochastic states without changing their $\ell^1$-norm. In order for the components of $\rho^i(t)$ to remain non-negative, the transition rate matrix $\Q$ should have non-negative off-diagonal elements. The condition \eqref{infstoch} then fixes the diagonal elements as minus the sum of all non-diagonal elements in that column. So any $n_c \times n_c$-dimensional matrix $\Q$ is \emph{infinitesimally stochastic} if its elements satisfy $\Q_{\Y\X} \geq 0$ for $\X \neq \Y$ and
\begin{equation}
\Q_{\X\X} = - \sum_{\substack{\Y \in C \\  \Y \neq \X} } \Q_{\Y\X} \,.
\end{equation}
For two-dimensional matrices, a useful basis for the infinitesimally stochastic matrices is
\begin{equation}\label{h2c}
\hSI = \left( \begin{array}{cc} -1 & 0 \\ 1 & 0 \end{array} \right) \,, \qquad 
\hIS = \left( \begin{array}{cc} 0 & 1 \\ 0 & -1 \end{array} \right) \,, \qquad 
\end{equation}
In general, there are $n_c(n_c-1)$ linearly independent infinitesimally stochastic matrices and we define $\h_{\textsc{x} \to \textsc{y}}$ as the matrix with all zero elements except a minus one on the diagonal of the $\X$-th column and a plus one the $\Y$-th row of that column:
\begin{equation}\label{hXY}
\h_{\textsc{x} \to \textsc{y}} = \ket{\Y}\bra{\X} - \ket{\X}\bra{\X} \,.
\end{equation}
Using this basis, any transition rate matrices $\Q$ for the master equation \eqref{msteqn} can be written as a linear combination of $\hXY$, such that:
\begin{equation}\label{Hgenlin}
\Q = \sum_{\substack{\X,\Y \in C \\ \X \neq \Y}} \epsilon_{\X\Y} \hXY\, ,
\end{equation}
where  $\epsilon_{\X\Y} $ are non-negative transition rates. For any such $\Q$, the rate of change of the total probability is zero by \eqref{msteqn} and the individual probability vectors will remain normalized under the $\ell^1$-norm \eqref{L1norm}. Furthermore, each component of $\ket{\rho^i}$ will stay bounded between zero and one if:
\begin{equation}\label{epsilonconstraint}
\forall \A:  0 \leq \sum_{\X \neq \A} \epsilon_{\A \X} \leq 1\,.
\end{equation}
This bound on $\epsilon_{\A\X}$ signifies that the sum of transition rates out of any state $\A$ cannot exceed one\footnote{Note that even though individual transition rates are constrained by \eqref{epsilonconstraint}, ratios of them can be arbitrarily large. Sometimes, one of the $\epsilon_{\X\Y}$'s will be removed by rescaling time $t \to t / \epsilon_{\X\Y}$ to form a dimensionless time. In that case, ratios of the transition rates might appear explicitly in the infinitesimal generator, seemingly in violation of \eqref{epsilonconstraint}. This is however only artificial, since we can always scale back to the physical time for which \eqref{epsilonconstraint} holds}.

\subsection{The tensor product formulation}
\label{sec:tenprodform}

The state of the population is a $n_c^N$ dimensional vector, whose components correspond to the probability of finding the system in each of its $n_c^N$ possible microscopic configurations. This many-body probability vector is an element of the tensor product space of the individual vector spaces: 
\begin{equation}\label{rhogen}
\ket{\rho(t)} \in \cP_{\rm m.b.}^{(N)} = \cP_1^{n_c} \otimes \cP_2^{n_c} \otimes \ldots  \otimes \cP_N^{n_c}\,.
\end{equation}
The tensor product $ V \otimes W$ is a bilinear map from two vector spaces $V, W$ to the product space $V \otimes W$, which is a $\text{Dim}(V) \times \text{Dim}(W)$ dimensional vector space. For vectors in a specific basis, the result is a vector $v \otimes w$, whose elements are all possible combinations of the products of the elements of $v \in V$ and $w \in W$. Some basic properties about tensor products are recalled in appendix \ref{sec:appendixA}. 

When there are no dynamical correlations in the system, it suffices to specify only the state of each of the $N$ nodes. The state is then called \emph{separable}, analogously to quantum states. The separable state is given as the direct product of the probability vectors of all $N$ nodes in the network
\begin{equation}\label{rhopop}
\ket{\rho(t)} = \ket{\rho^1(t)} \otimes \ket{\rho^2(t)} \otimes \ldots \otimes \ket{\rho^N(t)} = \ket{\rho^i(t)}^{\otimes_N} \,.
\end{equation}
In general, interactions between individual nodes will correlate their states. It is then no longer possible to write the state of the population as a single separable state. The generic state \eqref{rhogen} can be expressed as a linear combination of basis states
\begin{equation}
\ket{\rho(t)} = \sum_{i_n \in C} \rho_{i_1 \ldots i_N}(t) \ket{ i_1 \ldots i_N} \,,
\end{equation}
where $\ket{i_1 \ldots i_N} \in C^N$ denote all possible deterministic states of the composite system. The normalization condition on $\ket{\rho(t)}$ implies that the components $ \rho_{i_1 \ldots i_N}(t) $ form a discrete probability distribution, where now the $\ell^1$-norm of $\ket{\rho(t)}$ is computed by generalizing the co-vector $\bra{1}$ to the tensor product of $N$ unit co-vectors
\begin{equation}
\bra{1} = \left(1  \;\ldots \; 1 \right)^{\otimes_N}\,.
\end{equation}
In practice, we will simply write $\bra{1}$ for any vector with all elements equal to one and let its dimensionality be clear from the context.  As a consequence of \eqref{L1norm}, the separable states $\ket{\rho(t)}$ in \eqref{rhopop} are normalized.

The composite vector $\ket{\rho(t)}$ also evolves in time by the master equation 
\begin{equation}\label{mastereqn}
\partial_t \ket{\rho(t)} = \Q \ket{\rho(t)} \,,
\end{equation}
but now $\Q$ is a $n_c^N \times n_c^N$-dimensional transition rate matrix. It is possible to construct $\Q$ as a tensor product of $N$ $n_c$-dimensional matrices. This is the main advantage of using tensor products, as it also acts on linear maps of vector spaces, such as matrices. For arbitrary $n_c$-dimensional matrices $\cO_1$ and $\cO_2$, the tensor product $\cO_1 \otimes \cO_2$ is a linear map on the product space and
\begin{equation}\label{linmaps}
(\cO_1 \otimes \cO_2) \left( \ket{\rho_1} \otimes \ket{\rho_2} \right) = \cO_1 \ket{\rho_1} \otimes \cO_2 \ket{\rho_2} \,.
\end{equation}
We can hence think of $\cO_1 \otimes \cO_2$ as applying a local operation $\cO_1$ at the first site and $\cO_2$ at the second site. This means that we can construct transition rate matrices $\Q$ for the many-body master equation \eqref{mastereqn} as bilocal operators, acting only on two nodes if they share an edge in the network
\begin{equation}\label{bilocalQ}
\Q = \sum_{i,j}^N \epsilon^{ij} A^{ij} \cO_1^i \cO_2^j + \sum_{i=1}^N \epsilon^i \cO_3^i \,.
\end{equation}
Here $\cO^i$ denotes a tensor product of $N-1$ identity matrices with the operator $ \cO$ inserted at site $i$, such that it can act locally on $\ket{\rho^i}$ within the tensor product. The product term $\cO_1^i \cO_2^j$ denotes the matrix multiplication of two such operators, which can be understood as a bilocal operator, simultaneously acting with $\cO_1$ at site $i$ and $\cO_2$ at site $j$, while all other sites are unaffected.

To obtain the transition rate matrices for any specific model, we should pick the operators $\cO_{1,2,3}$ according to rules specified by the compartmental model. In any case, the operators should be chosen such that $\ \cO_1 \otimes \cO_2$ and $\cO_3$ are infinitesimally stochastic. Fortunately, the property \eqref{linmaps} ensures that
\begin{equation}\label{tensprodinfstoch}
	\bra{1} (\cO \otimes \hXY) = \bra{1}\cO \otimes \bra{1}\hXY  = \bra{0}\,,
\end{equation}
for arbitrary $\cO$. Hence any tensor product involving at least one $\hXY$ is itself a matrix with vanishing column sums. If one further ensures that the off-diagonal elements are non-negative, then this product is infinitesimally stochastic. The particular operators which appear in \eqref{bilocalQ} will determine the microscopic interaction rules of the many-body Markov chain. In the next section we will discuss this in detail for the $\S\I$ model and generalize to other compartmental models on networks in section \ref{sec:compartmentals}.

\section{The SI transition}
\label{sec:Logisticgate}

Any compartmental model of epidemiology has at its core an $\textsc{s} \to \textsc{i}$ transition. It is the transition responsible for the transmission of the infection and it depends on the status of two individuals. A susceptible and infected meet, interact, and with some probability, the susceptible becomes infected.
The deterministic rate equation for the fraction of infected in the population $i(t)$ is the logistic differential equation, introduced by Verhulst in 1838 to model population growth \cite{verhulst1838} and studied in the epidemiological context in \cite{bailey1950simple}. This equation describes the macroscopic system at the level of the (homogeneously mixed) population
\begin{equation}\label{logisticeqn}
\partial_t i(t) = \alpha s(t) i(t) = \alpha \left(1- i(t)\right)i(t) \,.
\end{equation}
Here we have used that the fraction of susceptible $s(t)$ satisfies $s(t) = 1 - i(t)$ and we have defined $\alpha$ as the macroscopic transmission rate. 

The non-linear dynamics of \eqref{logisticeqn} is a result of the bilocal infection transition of the microscopic model \eqref{bilocalQ}. In this case, 
the transition rate matrix $\Q_{\S\I}$ for the $\StoI$ transition can be constructed by performing the following local operations on each pair of nodes $(i,j)$ sharing an edge in the network. At node $i$ we check whether it is infected by projecting with $P_\I$. At the second node, we act with the infinitesimally stochastic matrix $\hSI$. This matrix takes away from the susceptible component and adds to the infected component of the node. It does so only if $j$ is susceptible, as these non-zero entries appear in the first column of the matrix. In effect this means that node $i$ infects node $j$, only if $i$ has non-zero infected probability and $j$ has non-zero susceptible probability. Hence the transition rate matrix for the microscopic $\S\I$ model is:
\begin{equation}\label{HSI}
	\Q_{\S \I}(\tau, A^{ij})  = \tau \sum_{\substack{i,j =1 \\ i \neq j}}^N A^{ij} P_\I^i \, \hSI^j 
\end{equation}
Here we have defined $\tau$ as the microscopic transmission rate. In principle, each microscopic transition could appear with a unique transition rate $\tau_{ij}$, characterizing the fact that some interactions have a higher/lower change of transmission than others. However, since the $\StoI$ transition is the only two-body transition in most compartmental models, this effect can be modeled by choosing the highest transmission rate to equal $\tau$ and normalizing the corresponding entry of $A^{ij}$ to one. All other entries of the adjacency matrix are then positive real numbers between 0 and 1. If the network $A^{ij}$ is undirected, $A^{ij}$ will be symmetric and the infection can travel both ways along the edge. For directed graphs, the infection will spread only from $i$ to $j$. When we discuss exact solutions for this $\S\I$ system in section \ref{sec:exact}, we will assume undirected graphs and set all non-zero elements of the adjacency matrix to one.

\subsection{Two-node transition matrix as a stochastic logic gate}
To see how the $\StoI$ transition works in detail, lets first look at the most simple non-trivial graph, two nodes $a$ and $b$ connected by an edge. We will suppose for simplicity that $A^{ab} = 1 = A^{ba}$ and set the transmission rate $\tau=1$, which can be done by scaling time as $t \to t/\tau$. The two-node transition rate matrix is explicitly
\begin{align}
	\Q_{\S\I}^{N=2} & = (P_\I \otimes \mathbb{1}) (\mathbb{1} \otimes \hSI) + (\mathbb{1} \otimes P_\I ) (  \hSI \otimes \mathbb{1}) \nonumber \\
	&
	= \left( \begin{array}{cc}
	0 & 0 \\
	0 & 1
	\end{array}\right) \otimes \left( \begin{array}{cc} -1 & 0 \\ 1 & 0 \end{array} \right) +
	  \left( \begin{array}{cc}
	-1 & 0 \\
	1 & 0
	\end{array}\right) \otimes \left( \begin{array}{cc} 0 & 0 \\ 0 & 1 \end{array} \right)
	\nonumber \\
	& = 
	  \left( \begin{array}{cccc}
	0 & 0 & 0 & 0 \\
	0 & -1 & 0 & 0 \\
	0 & 0 & -1 & 0 \\
	0 & 1 & 1 & 0
	\end{array} \right) \,.
\end{align}
Since this matrix satisfies $\Q_{\S\I}^2 = - \Q_{\S\I}$, the matrix exponent is easily computed and gives
\begin{equation}
	\hat S(t) = \exp \left( \Q_{\S\I}^{N=2} t \right) = \left( \begin{array}{cccc}
	1 & 0 & 0 & 0 \\
	0 & e^{- t} & 0 & 0 \\
	0 & 0 & e^{- t} & 0 \\
	0 & 1- e^{- t} & 1- e^{- t} & 1
\end{array} \right) \,,
\end{equation}
This matrix contains all conditional probabilities $P(\Y_t,t|\Y_0,0)$, where the columns denote the initial microscopic state $\Y_0 \in \{\S,\I\}^2$ and the rows denote the final state $\Y_t$. It is interesting to think of the transition rate matrix $\Q_{\S\I}$ as a stochastic logic gate. 
\begin{equation}\label{stochgate}
\begin{array}{l} \includegraphics[width=2cm]{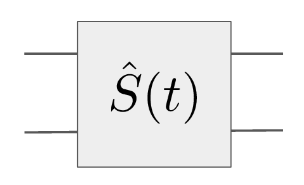} \end{array}
\end{equation}
The two open edges on the left signify that this operator takes as input the product state of two individual probability vectors and the open edges on the right signify that the output is again a product of two individual vectors. Depending on the initial conditions, this stochastic gate performs the following operation. If both input vectors are in the same state, i.e. both susceptible or both infected, nothing happens. But if one of the two inputs is infected and the other susceptible, then the latter will become infected with transmissibility $1-e^{-t}$. This can be computed from the expectation value of the susceptible node after interacting. Suppose that we start with $a$ susceptible and $b$ infected, such that
\begin{equation}\label{initcon}
\ket{\rho(0)} = \ket{\S} \otimes \ket{\I} \equiv \ket{\S\I} \,,
\end{equation}
Then, the expectation value for the infected probability of $a$ is computed as
\begin{equation}
\vev{\I^a(t)} = \bra{1} P_\I^a \hat S(t) \ket{\S\I}  = 1 - e^{- t} \,.
\end{equation}
Here $P_\I^a$ is defined as acting with $P_\I$ on site $a$ and the identity on site $b$, i.e. $P_\I^a = P_\I \otimes \mathbb{1}$.

In general, the input vectors need not be deterministic (as in, either $\ket{\S}$ or $\ket{\I}$). They do not even need to be separable states (as in \eqref{rhopop}). The approach is similar to constructing quantum logic gates in the quantum circuit model \cite{nielsen2002quantum}. The crucial difference, however, is that quantum gates act as unitary operators on complex wave functions, while here $\hat{S}(t)$ is a stochastic operator acting on non-negative and real probability vectors.

\subsection{Correlations as non-separable states}
When acting with the stochastic logic gate \eqref{stochgate} on a direct product state 
\begin{equation}
	\ket{\rho^{ab}} = \ket{\rho^a} \otimes \ket{\rho^b}\,,
\end{equation}
the resulting state is explicitly
\begin{equation}\label{stochentangled}
\hat S(t) \ket{\rho^{ab}} = \left( \begin{array}{c}
\rho_\S^a \rho_\S^b \\
e^{-t}\rho_\S^a \rho_\I^b \\
e^{-t}\rho_\I^a \rho_\S^b \\
\rho_\I^a \rho_\I^b + (1-e^{-t}) (\rho^a_\S \rho^b_\I + \rho^a_\I \rho^b_\S)
\end{array}\right) \,.
\end{equation}
This two-node state can not be written as a separable state \eqref{rhopop}, which implies that the nodes $a$ and $b$ are \textit{correlated}. The composite probability distribution does not factorize into the product of the component probability distributions. If these vectors would be wave functions in a Hilbert space, the composite state would be called entangled. Mathematically, the definitions of quantum entanglement and classical correlation both rely on the same inability to factorize a state into a direct tensor product of vectors. It is only the nature of the vector space which distinguishes entanglement (Hilbert spaces) from classical correlation (probability spaces).

The large $t$ limit of the state \eqref{stochentangled} gives
\begin{equation}\label{stochBell}
\rho^{ab}(t \to \infty) = \left( \begin{array}{c}
\rho_\S^a \rho_\S^b \\
0 \\
0 \\
\rho_\I^a \rho_\I^b + \rho^a_\S \rho^b_\I + \rho^a_\I \rho^b_\S
\end{array}\right) \,.
\end{equation}
This shows there are two steady states in the system: the completely susceptible state $\ket{\S\S}$ and the completely infected state $\ket{\I\I}$. All states other than $\ket{\S\S}$ will eventually reach the completely infected state. If the initial conditions $\rho^i_\X$ for $i =a ,b$ and $\X = \S, \I$ are unknown, performing a measurement on one node of the late-time state \eqref{stochBell} would give the state of the other node with absolute certainty.
However, it is not true that from the late time state the initial conditions can be recovered.

\subsection{Node-based mean-field approximation}
\label{sec:nodebased}

We continue with discussing how the node-based dynamics of the $\S\I$ model on any graph can be recovered from the transition rate matrices \eqref{HSI}. To this end, we compute the rate of change of the expected infected probability of an arbitrary node $i$ in the network
\begin{equation}
\partial_t \vev{\I^i(t)} =  \partial_t  \bra{1} P_\I^i \ket{\rho(t)}  \,,
\end{equation}
where again, $P_\I^i$ is defined as acting with $P_\I$ only on node $i$ and acting with the identity operator on all other nodes.
By the master equation, this becomes
\begin{equation}
\partial_t \vev{\I^i(t)} =  \tau \bra{1} P_\I^i    \sum_{\substack{j,k =1 \\ j \neq k}}^N A^{jk} P_\I^j \, \hSI^k    \ket{\rho(t)} 
\end{equation}
The property $\bra{1} \hXY = \bra{0}$ ensures that the only contribution comes from the terms where the projection $P_\I^i$ acts on the same site as $\hSI^k$. At that node, acting first with $\hSI$ and then with $P_\I$ is equivalent to acting with the matrix product $P_\I \, \hSI = \tilde{P}_\S$, giving:
\begin{equation}
\partial_t \vev{\I^i(t)} =  \tau \sum_{\substack{j =1 \\ j\neq i}}^N  A^{ji}  \bra{1}   P_\I^j \, \tilde P_\S^i    \ket{\rho(t)} \,,
\end{equation}
where $\tilde P_\S$ is defined in \eqref{Ptilde}. For separable states \eqref{rhopop} the projection operators return the corresponding components from the individual probability vectors. This gives the individual based mean-field approximation (IBMF) for the $\S\I$ model on generic graphs (see for instance chapter 17 of \cite{newman2010networks} or \cite{van2008virus} with zero recovery)
\begin{equation}\label{SIibmf}
\partial_t \rho_\I^i (t) = \tau \sum_{\substack{j =1 \\ j \neq i}}^N  A^{ji} \rho_\S^i(t) \rho_\I^j(t) =  \tau (1- \rho_\I^i(t))\sum_{\substack{j =1 \\ j \neq i}}^N  A^{ji}  \rho_\I^j(t) \,. 
\end{equation}
If the state $\ket{\rho(t)}$ contains dynamical correlations, we know from the previous section that it cannot be written as a separable state \eqref{rhopop}. In that case, we define the expectation value of having an $\S\I$ pair in the network as
\begin{equation}
\vev{\S^i \I^j (t) } = \bra{1} \tilde P^i_\S \, P_\I^j \ket{\rho(t)} \,.
\end{equation}
Its evolution in time is computed by using the master equation once more and it depends on the presence of triples
\begin{align}
\partial_t \vev{\S^i \I^j (t) } = & \, \bra{1} \tilde P^i_\S \, P_\I^j \Q_{\S\I} \ket{\rho(t)} \\
 = & \,   \tau  \sum_{\substack{k=1 \\ k \neq i,j}}^N  A^{kj} \vev{\I^k \S^j \S^i(t) } - \tau  \sum_{\substack{k=1 \\ k \neq i,j}}^N  A^{ki} \vev{\I^k \S^i \I^j(t) }  \nonumber  - \tau A^{ji} \vev{\S^i \I^j (t) } 
\end{align}
with:
\begin{subequations}\label{triples}
\begin{align}
\vev{\I^k \S^i \I^j(t) } & =  \bra{1} P_\I^k \, \tilde P_\S^i  \, P_\I^j \ket{\rho(t)} \,, \\
\vev{\I^k \S^j \S^i(t) } & =  \bra{1} P_\I^k \, \tilde P_\S^j  \, \tilde P_\S^i \ket{\rho(t)} \,. 
\end{align}
\end{subequations}
In turn, the rate equation for the expectation values of the triples \eqref{triples} will depend on the quartic terms, as well as on the triples. This interdependence of expectation values on terms involving more nodes will continue until the network size $N$ is reached. In practice, often a mean-field type approximation is made by `closing' the set of differential equations at some order $n$ \cite{kiss2017mathematics}. This can be done by assuming that the expectation value involving $n$ nodes are not statistically independent from those at a lower order, stopping the chain of interdependence. Of these, pair-approximations are most popular, see for instance \cite{ben1992mean,rand1999correlation,keeling1999effects}. In principle, though, the full set of rate equations can be derived using only the transition rate matrix $\Q_{\S\I}$ in \eqref{HSI} and the process outlined here.

We close this section by discussing how homogeneous mixing, combined with the mean-field approximation give the Verhulst logistic equation \eqref{logisticeqn} in the limit of large population size. First we define the total fraction of infected in the population as
\begin{equation}
i(t) = \frac{1}{N} \sum_{i=1}^N \vev{\I^i} =  \frac{1}{N} \sum_{i=1}^N \bra{1} P_\I^i \ket{\rho}\,,
\end{equation}
and likewise for $s(t)$. Using the mean-field approximation \eqref{rhopop} for $\ket{\rho}$ and the master equation leads to
\begin{equation}\label{meanfielit}
\partial_t i(t) = \frac{\tau}{N} \sum_{ \substack{i,j=1 \\ i\neq j}}^N A^{ij} \rho_\S^i(t) \rho^j_\I (t) \,.
\end{equation}
The homogeneous mixing approximation is made by taking each node to have the same average probability of interacting with any other node, so we may replace $\tau A^{ij}$ by an effective average transmission rate $\vev{\tau} $. Using the definitions of $i(t)$ and $s(t)$ leads to
\begin{equation}
\partial_t i(t) = \vev{\tau} N s(t) i(t) - \frac{\vev{\tau}}{N} \sum_{i=1}^N \rho_{\I}^i(t) \rho_\S^i (t)\,.
\end{equation}
The second term here is to account for the fact that the sum in \eqref{meanfielit} does not cover $i = j$. To recover \eqref{logisticeqn}, we should take the limit of large $N$, but keep $\vev{\tau} N = \alpha$ fixed. In that limit the last term is sub-leading and hence drops out. The macroscopic equations \eqref{logisticeqn} hence contain implicitly three approximations: homogeneous mixing, a mean-field approximation and a large $N$ limit. Within the framework described in this section, it is possible to relax all of these approximations in the microscopic model.

\section{Exact transition rate matrices for compartmental models}
\label{sec:compartmentals}

Although the bilocal $\StoI$ transition is an essential element of any compartmental model of epidemiology, its dynamics is limited because it only allows for the transmission of the infection. In more realistic models, individuals recover and, depending on the nature of the infection, could become immune to the pathogen. In order to account for this, additional transitions and compartments can be introduced. The increase in the number of parameters of the model typically leads to appearance of an epidemic threshold above which a non-zero fraction of the population will become infected.
Two benchmark models which allow for recovery are the $\SIS$ model \cite{kermack1932contributions} and the $\SIR$ model \cite{kermack1927}. Further refinements can be made by including, for instance, a class for exposed individuals which are not yet infectious (\textsc{e}), modeling the effect of an incubation period. 

In this section, we will describe how to construct the exact transition rate matrices for compartmental models with $n_c$ compartments and a generic number of transitions on arbitrary static networks. We will then provide as explicit examples some of the most widely used models in the literature and discuss their  main features briefly. Finally, this section is concluded by providing some ideas on further generalizations.

\subsection{General methodology}
\label{sec:genmet}
The generic compartmental model of epidemiology comes equipped with a specification of the compartments and the transitions between them. From the bilocal form of the infinitesimal generator \eqref{bilocalQ} we see that we should specify two different types of transitions, the bilocal interactions which depend on the neighbors state, and the local operators which act only on single nodes.
Each transition comes with its own transition rate, which gives the number of independent parameters in the model. The interaction transitions should also come with a specification of a \textit{control compartment}, which specifies the state of the neighbor which induces the transition. In the case of the $\StoI$ transition this was the $\I$ compartment, as only an infected individual can infect its susceptible neighbors. In general, though, the control compartment need not be equal to any of the compartments which the neighbor transitions into or out of.

Graphically, we denote the compartments as labeled boxes and transitions as arrows connecting the boxes. The two types of transitions are distinguished by two kinds of arrows, labeled by the transition rate and the control compartment in the case of the interaction transition:
\begin{equation}
\begin{array}{rl}
   \begin{tikzpicture}[decoration={ snake,  pre length=1pt,post length=3pt}]
]
\node[]  (1) at (0,0)    {};
\node[]  (2) at (1.5,0) {};

\path[draw=black,decorate,->] (1) -- node[above] {$\gamma$} (2);
\end{tikzpicture}
&: \quad \text{local transition with rate } \gamma
\\
\begin{tikzpicture}[decoration={ snake,  pre length=1pt,post length=3pt}]
]
\node[]  (1) at (0,0)    {};
\node[]  (2) at (1.5,0) {};

\path[draw=black,->] (1) -- node[above] {$\tau, \X$}  (2);
\end{tikzpicture}
& : \quad \text{bilocal interaction with rate } \tau
\\
& \quad \; \; \text{and control compartment } \X 
\end{array}
\end{equation}
The transition rate matrices are then constructed as follows:
\begin{itemize}
	\item For each local transition connecting compartments $\X$ and $\Y$ with transition rate $\gamma$, add a term $\hXY^i$ (defined in \eqref{hXY}), multiplied by the transition rate $\gamma$ for each node $i$ in the network, i.e.:
	\begin{equation}\label{intrinsictransition}
	\begin{array}{r}
	\begin{tikzpicture}[decoration={ snake,  pre length=1pt,post length=3pt}]
	\begin{scope}
	\node[shape=rectangle,draw=black,fill=green!20,minimum size=.8cm] (A) at (0,0) {$\X$};
	\node[shape=rectangle,draw=black,fill=yellow!20,minimum size=.8cm] (B) at (2,0) {$\Y$};
	\end{scope}
	
	\path[draw=black,decorate,->] (A) -- node[above] {$\gamma$} (B) ;
	\end{tikzpicture} 
	\end{array}
	: \quad \text{add} \;\; \gamma \sum_{i=1}^N \hXY^i\,.
	\end{equation} 
	\item For each interaction transition connecting compartments $\X$ and $\Y$ with transition rate $\tau$ and control compartment $\textsc{z}$, act with $P_{\textsc{z}}^i$ on one node and $\hXY^j$ on the other node of each connected pair in the network, i.e.:
	\begin{equation}\label{interactiontransition}
	\begin{array}{r}
	\begin{tikzpicture}[decoration={ snake,  pre length=1pt,post length=3pt}]
	\begin{scope}
	\node[shape=rectangle,draw=black,fill=green!20,minimum size=.8cm] (A) at (0,0) {$\X$};
	\node[shape=rectangle,draw=black,fill=yellow!20,minimum size=.8cm] (B) at (2,0) {$\Y$};
	\end{scope}
	
	\path[draw=black,->] (A) -- node[above] {$\tau, \textsc{z}$} (B) ;
	\end{tikzpicture} 
	\end{array}
	: \quad \text{add} \;\; \tau \sum_{\substack{i,j=1 \\ i \neq j}}^N A^{ij} P_{\textsc{z}}^i \, \hXY^j\,.
	\end{equation} 
\end{itemize}
The sum of all transitions in the compartmental model then gives the exact transition rate matrix for that model. In the following subsections, we will discuss several explicit examples of this general construction.

\subsection{The $\SIS$ model}
\label{sec:SIS}

The $\SIS$ compartmental model of \cite{kermack1932contributions} extends the $\StoI$ transition of the logistic growth model by introducing a recovery transition. By a Poisson process, infected individuals are recovered with rate $\gamma$. In the $\SIS$ model, the recovery process does not infer immunity to the host, so the recovered individual is moved back to the susceptible compartment. The deterministic rate equations for the macroscopic $\SIS$ system in terms of the fraction of infected $i(t)$ and susceptible $s(t)$ in the population are
\begin{subequations}
\label{sis}
\begin{align}
\partial_t s(t) & = - \alpha s(t) i(t) + \beta i(t) \,, \\
\partial_t i(t) & =  \alpha s(t) i(t) - \beta i(t) \,. \label{isis}
\end{align}
\end{subequations}
It is well-known that this model has an equilibrium phase transition from an absorbing to an endemic phase at a critical value of $ R_0 = \frac{\alpha}{\beta}  $. This can easily be seen by computing the steady state solutions at late time $i(\infty)$. There are two solutions
\begin{equation}
	i(\infty) = 0 \,, \qquad \text{or:} \quad i(\infty) = 1 - \frac{1}{R_0} \,.
\end{equation}
The latter solution is only positive in the regime $R_0>1$. The fact that the macroscopic system will always flow to the latter solution whenever $R_0>1$ is illustrated best by defining the infected potential $V(i(t))$. We may write \eqref{isis} using $s(t) = 1- i(t)$ as
\begin{equation}
\partial_t i(t) = - \frac{\partial}{\partial i(t)} V(i(t)) \,,
\end{equation}
with
\begin{equation}
V(i(t)) = \frac{\alpha}{3} i(t)^3 + \frac{\beta-\alpha}{2}i(t)^2\,.
\end{equation}
For $R_0<1$, this potential only has a single minimum in the range $0 \leq i(t) \leq 1$: the saddle-point at $i = 0$. However, when $R_0>1$, another minima of $V(i)$ appears in the range of interest, namely at $i = 1- \frac{1}{R_0}$. The value of the potential at this minima is always negative and so any perturbation away from $i =0$ will flow to this minimum. Hence, the macroscopic $\SIS$ system \eqref{sis} contains a stable endemic equilibrium with non-zero fraction of infected above the epidemic threshold. 

We can obtain the microscopic description of the $\SIS$ system from the diagrammatic representation of this compartmental model:
\begin{equation}\nonumber
\begin{array}{r}
\begin{tikzpicture}[decoration={ snake,  pre length=5pt,post length=5pt}]
\begin{scope}
\node[shape=rectangle,draw=black,fill=blue!20,minimum size=.8cm] (A) at (0,0) {$\S$};
\node[shape=rectangle,draw=black,fill=red!20,minimum size=.8cm] (B) at (2,0) {$\I$};
\end{scope}

\path[draw=black,->] (A) -- node[above] {$\tau, \I$} (B);
\path[draw=black,->] (B) edge[bend left=90,decorate] node[above] {$\gamma$} (A) ;
\end{tikzpicture} 
\end{array}
\end{equation}
The rules of the last subsection allow to write down the exact microscopic transition rate matrix immediately as:
\begin{align}\label{Hsis}
\Q_{\SIS}(\tau,\gamma,A^{ij}) & = \tau \sum_{\substack{i,j =1 \\ i \neq j}}^N A^{ij} P_\I^i \, \hSI^j + \gamma \sum_{i=1}^{N} \hIS^i
\end{align}
For static networks, the system is hence solved by computing $\exp({\Q_{\SIS} t} )$.
In practice, though, computing this matrix exponential explicitly for a sizable network is a formidable task. The discussion section \ref{sec:conclusion} will explore some techniques which can be utilized in extracting useful information from the matrix exponential. Here, we will discuss only the two-node $\SIS$ system in more detail.

\paragraph{Two-node $\SIS$ model}

The transition rate matrix for the connected two-node $\SIS$ system is:
\begin{align}\label{HsisN2}
\Q_{\SIS}^{N=2} & = \left(
\begin{array}{cccc}
0 & \gamma & \gamma & 0 \\
0 & -\tau-\gamma & 0 & \gamma \\
0 & 0 & -\tau-\gamma & \gamma \\
0 & \tau & \tau & -2 \gamma \\
\end{array}
\right) \,.
\end{align}
From this it is possible to compute the sum of the infected probabilities $\vev{\I(t)}$ as
\begin{equation}
\vev{\I(t)}  = \sum_{i=1}^2 \bra{1} P_I^i e^{ \Q_{\SIS}^{N=2} t } \ket{\rho(0)} \,.
\end{equation}
If we start with the initial conditions \eqref{initcon} with one node infected and the other susceptible, we find explicitly:
\begin{equation}\label{Itsis}
\vev{\I(t)} = \frac{e^{-\frac{r+3}{2} t} }{\sqrt{\alpha}} \left( \sqrt{\alpha} \cosh \left(\frac12 \sqrt{\alpha} t\right) + (1+3r) \sinh \left( \frac12 \sqrt{\alpha} t\right) \right)\,,
\end{equation}
with $\alpha = 1+ r(6+r)$ and $r= \tau/\gamma$.
\begin{figure}[t]
	\centering
	\includegraphics[width=.5 \textwidth]{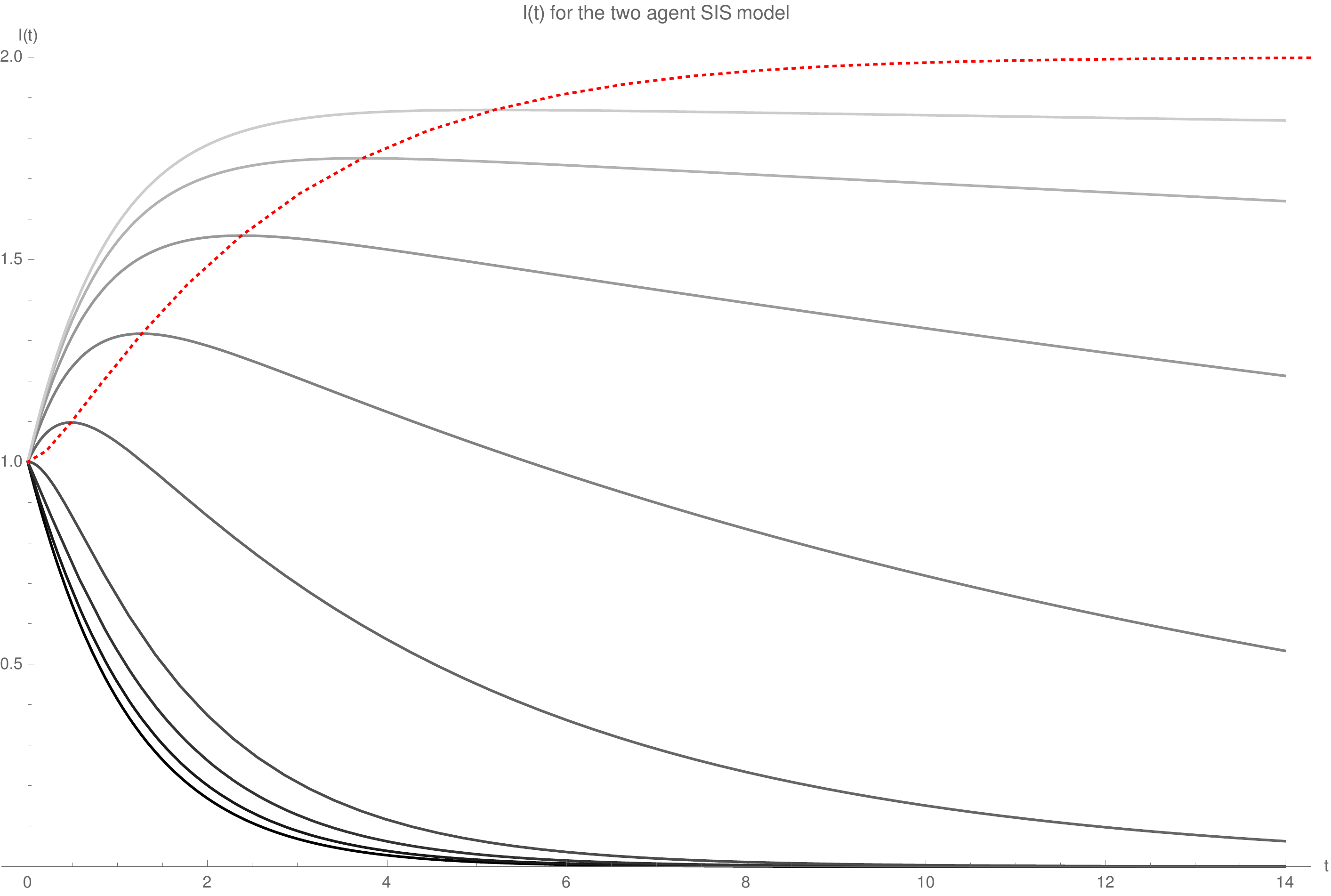}
	\caption{The total infected probability in the two node $\SIS$ system for initial conditions where a single node is infected and the other susceptible. The plots are drawn for $r = 2^n$ with $n$ increasing from $-3$ (darkest) to $5$ (lightest). The red line indicate the maxima of $\vev{\I(t)}$ as a function of $r$.}
	\label{fig:SIS}
\end{figure}
This function is plotted in figure \ref{fig:SIS} for various values of $r$. We see explicitly that the steady state solution for the exact two-node system is the infectionless state $\ket{\S\S}$, as can be verified by taking limit $t \to \infty$ of \eqref{Itsis}. In this sense, the equilibrium phase transition to an endemic state has disappeared in the microscopic model, because there is only a single steady state: the absorbing state $\ket{\S\S}$ \cite{ganesh2005effect,van2008virus}. However, figure \ref{fig:SIS} shows that for $r>1$ the relaxation towards equilibrium shows a qualitative difference compared to the $r<1$ curves. For $r>1$, the total infected curve first passes through a maximum, before relaxing towards the steady state $\ket{\S\S}$. If $r$ is significantly bigger that $r_c=1$, the relaxation towards equilibrium is slow and the maxima flattens to become a meta-stable state with high infection probability at intermediate timescales. The system hence possesses an out-of-equilibrium phase transition, where depending on the value of the control parameter $r$, a non-trivial maximum of the order parameter $\vev{\I(t)}$ appears. This maxima has been plotted for $r>1$ as the red line in figure \ref{fig:SIS}.

To see this explicitly, we look for states which satisfy
\begin{equation}\label{Isis_steadystate}
\partial_t\vev{\I^i(t)} = \bra{1} P_\I^i  \Q_{\SIS}^{N=2} \ket{\rho} =0 \,,
\end{equation}
for all $i$. If we assume that $\ket{\rho}$ is separable, we can write $\ket{\rho} = \ket{\rho^a} \otimes \ket{\rho^b}$ and find two solutions to \eqref{Isis_steadystate}:
\begin{equation}
\ket{\rho}_1 = \ket{\S\S} \,, \qquad \text{and:} \quad  \ket{\rho}_2 = \left( \begin{array}{c}
\frac{1}{r} \\
1 - \frac{1}{r}
\end{array}\right) \otimes
\left( \begin{array}{c}
\frac{1}{r} \\
1 - \frac{1}{r}
\end{array}\right)
\end{equation}
The first solution is the absorbing state, valid for all values of $r$. The second solution only represents a sensible probability distribution when $r> r_c =1$ and it corresponds to the individual based mean-field solution of the $\SIS$ system on the complete graph with $N=2$ (see, for instance, \cite{van2008virus}).

We could ask whether relaxing the mean-field approximation would lead to new solutions of \eqref{Isis_steadystate}. By parameterizing all possible states of $\ket{\rho}$ explicitly, it is possible to find a continuous family of solutions to \eqref{Isis_steadystate}, parameterized by a real parameter $x$ constrained to the interval $0 \leq x \leq \frac{1}{1+r}$:
\begin{equation}\label{rhox}
\ket{\rho(x)} = \left(1 - (r+1)x\right)\ket{\S\S} + x \left(\ket{\S\I} + \ket{\I\S} \right) + x(r-1) \ket{\I\I} \,.
\end{equation}
Requiring all elements of $\ket{\rho(x)}$ to lie in the range $\mathbb{R}^{[0;1]}$ restricts $r$ to be at or above the critical value $r\geq 1$ (assuming $x \neq 0$). So even though including non-trivial correlations in the state leads to a continuous family of solutions to \eqref{Isis_steadystate}, all non-trivial solutions require $r\geq 1$ and hence the epidemic threshold of the mean-field approximation persists.

Note that for $x= \frac{r-1}{r^2}$ the solution \eqref{rhox} reproduces the mean-field solution $\ket{\rho}_2$. In addition, for any $x \neq 0$, the second time derivative $\partial_t^2 \vev{\I^i(t)}$ is always negative when $r\geq 1$, signaling the extremum is unstable. This is, of course, in line with the interpretation that it corresponds to a maximum of infected probability.

\paragraph{Mean-field approximation}
The individual-based mean field theory is recovered along similar lines as discussed for the $\S\I$ model in section \ref{sec:nodebased}. We take $\ket{\rho}$ to be a separable state \eqref{rhopop} and compute the rate of change of the infected probability of individual nodes using the master equation and the transition rate matrix \eqref{Hsis}. The result gives
\begin{equation}
\partial_t \vev{\I^i(t)} = \partial_t \rho_\I^i(t) = \tau \sum_{j\neq i} A^{ij} \rho_\S^i(t) \rho_\I^j (t) - \gamma \rho_\I^i(t) \,.
\end{equation}
This corresponds to the mean-field approximation closed at the level of individuals, or NIMFA \cite{van2008virus}. Pair-based mean-field approximations can be recovered along the lines sketched in section \ref{sec:nodebased} and reproduce known results in the literature (see, for instance, \cite{kiss2017mathematics}). The limit to the macroscopic system \eqref{sis} is recovered analogously to the limit to the logistic equation discussed at the end of section \ref{sec:nodebased}.

\subsection{The $\SIR$ model}
\label{sec:SIR}

The $\SIR$ model, first introduced by Kermack and McKendrick in 1927 \cite{kermack1927}, remains a benchmark in modern epidemiological modeling. Instead of having individuals recover back into the susceptible compartment, a third compartment $\R$ for recovered (or removed) is introduced. It models the situation where once infected, the individual is no longer susceptible to a second infection, either by acquiring immunity after the recovery process, or because they had failed to recover. 
The deterministic equations for the fraction of susceptible $s(t)$, infected $i(t)$  and recovered $r(t)$ in the population are
\begin{subequations}\label{sir}
\begin{align}
\partial_t s(t) & = - \alpha s(t) i(t) \,,\\
\partial_t i(t) & =  \alpha s(t) i(t) - \beta i(t)\,,\label{sirinf} \\
\partial_t r(t) & = \beta i(t) \,.
\end{align}
\end{subequations}
In this case, the steady state solution requires $i(\infty) = 0$, but it might pass through a non-trivial maximum before arriving there, as in the microscopic $\SIS$ system. A measure for this is having a non-zero fraction of the population in the recovered compartment at late times. To get an indication of this fraction, we notice that equation \eqref{sirinf} is also zero when $s(t) = \frac{\beta}{\alpha} = 1/R_0$, leading to a steady state at late times with $r(\infty) = 1 - \frac{1}{R_0}$. This is positive only when $R_0>1$.

The individual states of the microscopic model are now three dimensional vectors $\ket{\rho^i}= \rho_\S^i \ket{\S} + \rho_\I^i \ket{\I} + \rho_\R^i \ket{\R}$. The diagrammatic representation of the $\SIR$ compartmental model is:
\begin{equation}
\begin{array}{r}
\begin{tikzpicture}[decoration={ snake,  pre length=1pt,post length=2pt}]
\begin{scope}
\node[shape=rectangle,draw=black,fill=blue!20,minimum size=.8cm] (A) at (0,0) {$\S$};
\node[shape=rectangle,draw=black,fill=red!20,minimum size=.8cm] (B) at (2,0) {$\I$};
\node[shape=rectangle,draw=black,fill=green!20,minimum size=.8cm] (C) at (4,0) {$\R$};
\end{scope}

\path[draw=black,->] (A) -- node[above] {$\tau, \I$} (B);
\path[draw=black,->] (B) edge[decorate] node[above] {$\gamma$} (C) ;
\end{tikzpicture} 
\end{array}
\end{equation}
The microscopic system is obtained by following the general instructions of section \ref{sec:genmet}.
\begin{align}\label{Hsir}
\Q_{\SIR}(\tau,\gamma,A^{ij}) & = \tau \sum_{\substack{i,j =1 \\ i \neq j}}^N A^{ij} P_\I^i \, \hSI^j + \gamma \sum_{i=1}^{N} \hIR^i
\end{align}
The transition rate matrix for $\S\I\R$ is composed out of lower-triangular matrices. This implies that the tensor product is also lower-diagonal and hence the diagonal elements of $\Q_{\SIR}$ contain its eigenvalues, which are all linear functions of $\tau$ and $\gamma$.

\begin{figure}[t]
	\centering
	\includegraphics[width=.5 \textwidth]{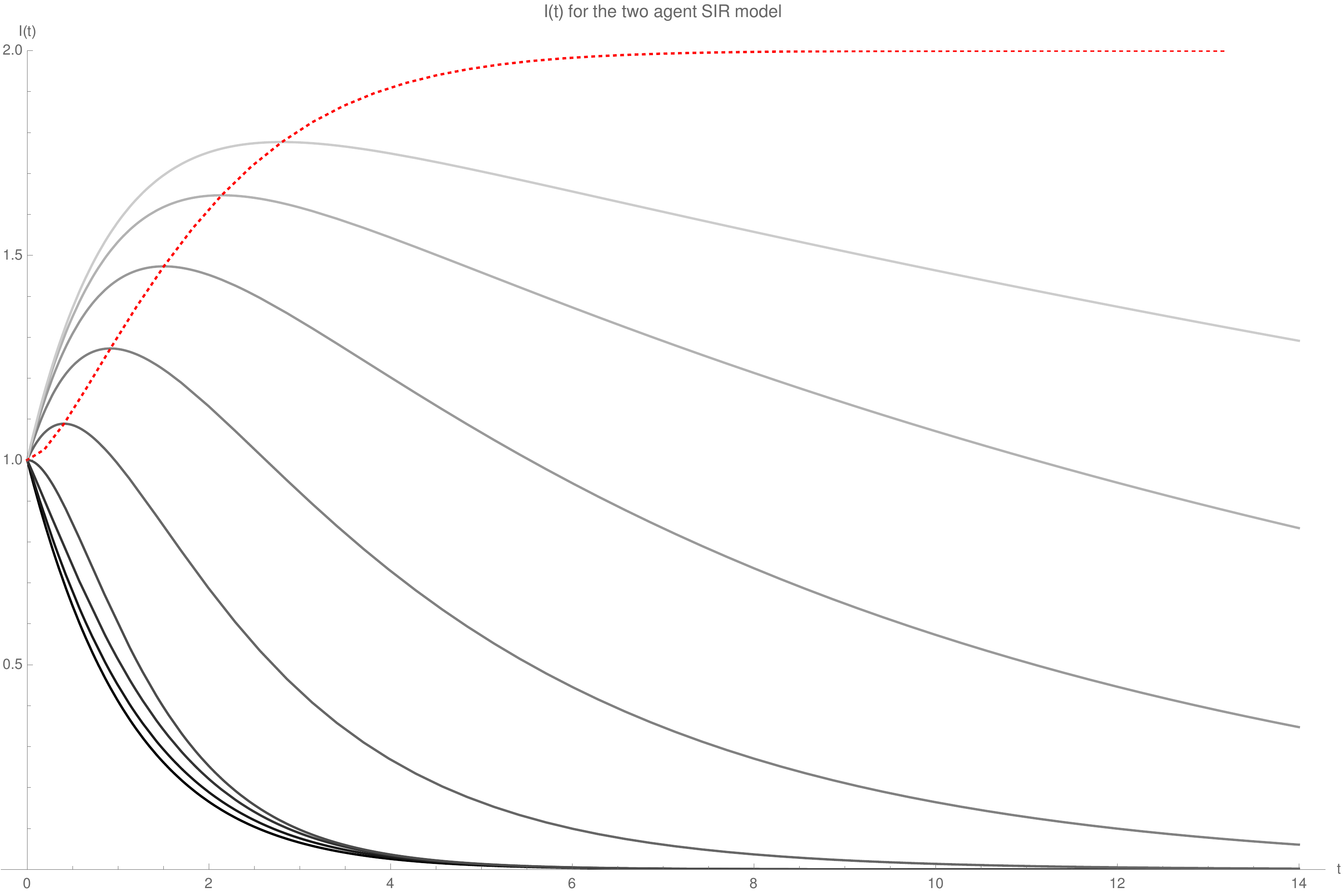}
	\caption{The total infected probability in the two agent $\SIR$ system for initial conditions where a single node is infected and the other susceptible. The plots are drawn for $r = 2^n$ with $n$ increasing from $-3$ (darkest) to $5$ (lightest). The red line indicate the maxima of $\vev{\I(t)}$ as a function of $r$.}
	\label{fig:SIR}
\end{figure}

Using \eqref{Hsir} for the two-node system, we can find the expectation value for the total number of infected as a function of time, starting from uncorrelated initial conditions $\ket{\rho(0)} = \ket{\rho^a} \otimes \ket{\rho^b}$:
\begin{equation}\label{Itsir}
	\vev{\I(t)} = e^{-(r+1)t} \left[ e^{rt}(\rho^a_\I + \rho^b_\I) + (e^{rt} - 1) (\rho_\S^a \rho_\I^b + \rho_\I^a \rho_\S^b) \right]\,,
\end{equation}
For initial conditions with one infected and one susceptible some curves for varying $r$ are drawn in figure \ref{fig:SIR}. 
Again, above $r=1$ a non-trivial maximum in \eqref{Itsir} appears. Before relaxing to any of the absorbing states (now any infectionless state is absorbing), the total infection probability peaks at (dimensionless) time $t_p$ given by:
\begin{equation}
t_p = \frac1r \log \left(\frac{1+r}{2} \right)\,.
\end{equation}
Only when $r>1$ is $t_p>0$, and so $r<1$ will have no non-trivial maximum in the positive time range. For $r=1$ the peak coincides with the initial conditions and for $r>1$ it is given as
\begin{equation}
\vev{\I(t_p)} =  2^{\frac1r} (r+1)^{-\frac1r} \left( 2 -  \frac{2}{r+1} \right) \,.
\end{equation}

From \eqref{Hsir}, existing mean-field approximations are obtained analogously to the $\S\I$ and $\SIS$ cases. The individual-based mean-field approximation for the simple two node system now leads to the uncorrelated steady state solution $\ket{\rho^i} = \frac1r \ket{\S} + x \ket{\I} + (1-\frac1r - x) \ket{\R}$ for all nodes $i$ and $0 \leq x \leq \frac{r-1}{r}$, with $r>1$. This solution reduces to the uncorrelated $\SIS$ steady state solution for the maximal infected probability $x = 1 - \frac1r$.

For generic networks, the exact value of the threshold where non-trivial extrema appear will depend on the network structure. This is much studied in the literature, due to the possible mapping of the $\SIR$ model to bond percolation. The epidemic threshold in that case is mapped to the percolation threshold above which a giant connected component forms. See for instance \cite{grassberger1983critical} and \cite{Newman2002} in the context of epidemic spreading on complex networks. Most of these methods, however, only discuss the static properties of the $\SIR$ systems, such as the final fraction of the population reached by the infection. The construction \eqref{Hsir} could be useful for an analytical analysis of the time-dependent behavior of the $\SIR$ system.

\subsection{Further generalizations}
\label{sec:further}

\paragraph{The $\SIR\S$ model}

A further generalization of the $\SIR$ model described above is the $\SIR\S$ model, where recovered individuals loose their immunity after some time. There can be several reasons for this. For instance, in the case of viral infections, the virus might mutate and produce a new strain, or the effect of removing from $\R$ and adding to $\S$ could be seen as mortality and birth rates, respectively \cite{liu2004spread}.
The compartmental diagram now contains an additional transition, connecting the $\R$ compartment with the $\S$ compartment:
\begin{equation}
\begin{array}{r}
\begin{tikzpicture}[decoration={ snake,  pre length=1pt,post length=2pt}]
\begin{scope}
\node[shape=rectangle,draw=black,fill=blue!20,minimum size=.8cm] (A) at (0,0) {$\S$};
\node[shape=rectangle,draw=black,fill=red!20,minimum size=.8cm] (B) at (2,0) {$\I$};
\node[shape=rectangle,draw=black,fill=green!20,minimum size=.8cm] (C) at (4,0) {$\R$};
\end{scope}

\path[draw=black,->] (A) -- node[above] {$\tau, \I$} (B);
\path[draw=black,->] (B) edge[decorate] node[above] {$\gamma$} (C) ;
\path[draw=black,->] (C) edge[bend left=60,decorate] node[above] {$\delta$} (A) ;
\end{tikzpicture} 
\end{array}
\end{equation}
The microscopic model is $\SIR$ with an additional local transition with rate $\delta$
\begin{equation}
\Q_{\SIR\S}  = \tau \sum_{\substack{i,j =1 \\ i \neq j}}^N A^{ij} P_\I^i \, \hSI^j + \gamma \sum_{i=1}^{N} \hIR^i
+ \delta \sum_{i=1}^N \h_{\R \to \S}^i \,.
\end{equation}
The mean-field approximations of the $\SIR\S$ model can be mapped to those of the $\SIS$ model \cite{bancal2010steady,liu2004spread} by a rescaling of the infected probabilities by $\frac{\delta}{\delta+1}$ and hence the critical properties are the same as in $\SIS$. 
However, when a time-delay is build into the $\SIR\S$ model by letting individuals be significantly delayed in the recovered class, oscillations can occur \cite{hethcote1981nonlinear}. Simulations on Watts-Strogatz small-world networks show an additional threshold above which a periodic steady state is observed \cite{kuperman2001small}. An interesting question for further research is whether the exact Markovian model also contains oscillatory steady states.

\paragraph{The $\S\E\I\R$ model}
Many influenza-like illnesses, including Covid-19, have an incubation (or latency) period in which an individual is exposed to the virus, but not yet infectious. To model this, a fourth compartment ($\E$ for exposed) can be introduced.
\begin{equation}
\begin{array}{r}
\begin{tikzpicture}[decoration={ snake,  pre length=1pt,post length=2pt}]
\begin{scope}
\node[shape=rectangle,draw=black,fill=blue!20,minimum size=.8cm] (A) at (0,0) {$\S$};
\node[shape=rectangle,draw=black,fill=purple!20,minimum size=.8cm] (B) at (2,0) {$\E$};
\node[shape=rectangle,draw=black,fill=red!20,minimum size=.8cm] (C) at (4,0) {$\I$};
\node[shape=rectangle,draw=black,fill=green!20,minimum size=.8cm] (D) at (6,0) {$\R$};
\end{scope}

\path[draw=black,->] (A) -- node[above] {$\tau, \I$} (B);
\path[draw=black,->] (B) edge[decorate] node[above] {$\gamma$} (C) ;
\path[draw=black,->] (C) edge[decorate] node[above] {$\delta$} (D) ;
\end{tikzpicture} 
\end{array}
\end{equation}
In this case, the transmission interaction is mediated by control compartment $\I$, but it transitions $\S \to \E$. In the microscopic model this is done by the following transition rate matrix:
\begin{equation}\label{HSEIR}
\Q_{\S\E\I\R}  = \tau \sum_{\substack{i,j =1 \\ i \neq j}}^N A^{ij} P_\I^i \, \h_{\S \to \E}^j + \gamma \sum_{i=1}^{N} \h_{\E \to \I}^i
+ \delta \sum_{i=1}^N \h_{\I \to \R}^i \,.
\end{equation}
As in the case of $\S\I$ and $\SIR$, the constituent matrices in \eqref{HSEIR} are lower-triangular, so also here the eigenvalues of the exact transition rate matrix are the elements of its diagonal, and hence they are linear functions of the transition rates. This property holds true for any compartmental model whose diagrammatic representation does not contain loops.

We could go on and discuss adding more transition and/or compartments, or generalize the framework to rumor spreading \cite{daley1965stochastic,zanette2002dynamics,nekovee2007theory}, voter-models \cite{sood2005voter,masuda2010heterogeneous} or other examples of social or population dynamics. Instead, we choose to focus in this work on the general framework, leaving specific applications for future work. 

Some limitations which have not been discussed up until this point is the inclusion of demographic data. Many of the more refined epidemiological models take input from known demographics such as age-distributions, as well as birth and mortality rates. The latter can be introduced by adding an additional compartment for diseased individuals to which all other compartments transition with mortality rate $\mu$. Birth rates can then be represented by a transition from the diseased compartment into the susceptible class. However, in view of the complexity of the problem, one might want to keep the number of compartments to a minimum and instead model birth and death rates by generalizing to dynamical adaptive networks, where new nodes can be added (birth) and existing ones removed (death). We will comment more on generalizations to adaptive networks in section \ref{sec:conclusion}. 

In general, though, the microscopic models allow for the possibility of assigning individual transition rates for each node. We could think of them as being picked from a distribution of transition rates. In that sense, age-related effects or pre-existing conditions could be included by picking slower recovery rates and higher mortality rates for some individuals. In fact, the assumption that all recovery rates are equal can be thought of as picking the individual recovery rates from a Poisson distribution, peaked sharply around the mean value \cite{Newman2002}. The total time spend in a compartment with an outgoing transition is then a random variable with an exponential distribution. To allow for non-Poissonian transmission and/or recovery, we would have to either pick the individual transition rates from a different distribution, or we could suppose that interactions are only turned on for a small time, and build in time-delays between interaction events.

\section{Perturbative expansion for SI outbreaks on graphs}
\label{sec:perturbative}

By constructing the transition rate matrix from bilocal operators on the product space of $N$ probability vector spaces, we can gain insight in the microscopic dynamics of the epidemic spreading process on networks. To illustrate this, we perturb the steady state by a small amount and study how the system responds. In this work we will be interested in tracking the relaxation towards equilibrium as a perturbative series in time. As the series is only expected to converge when the equilibrium is reached, we have to find a systematic way to compute contributions from arbitrarily high orders. This becomes possible because the perturbative expansion can be organized into sums over subgraph configurations. By viewing the transition rate matrix as a subgraph generating algorithm, we can track the type and multiplicities of the subgraphs contributing to the expansion. In this section, we will discuss how this works for the $\S\I$ model explicitly. We will show how this method can be used to find exact solutions by making use of the symmetry properties of the underlying graphs in the next section.

The $\S\I$ model has two steady state solutions: the completely infected population and the completely susceptible population. If we perturb around the completely infected population by flipping one node to susceptible, that node will quickly be infected and the system relaxes immediately to the steady state. On the other hand, if we flip one node to infected in the completely susceptible population, the infection will spread through the entire network until it has reached everybody and the state relaxes to the completely infected steady state. We say that the completely infected steady state is an attractor, because all perturbations around the $\ket{\S}^N$ steady state will flow there. The question is, can we compute this flow exactly using the transition rate matrices \eqref{HSI}?

To see how this process works, we start with initial conditions at $t=0$ of a single infected individual (patient 0) in a completely susceptible population:
\begin{equation}\label{patient0}
\ket{\rho^i(0)} = \ket{\I} \,, \quad \ket{\rho^j(0)} = \ket{\S} \,, \quad \forall j \neq i\,.
\end{equation}
The goal is to compute the time-dependent state $\ket{\rho(t)}$ such that we can use this to compute expectation values as a function of time. The solution to the master equation can be organized as
\begin{equation}\label{timeexp}
\ket{\rho(t)} = e^{t \Q_{\S\I}} \ket{\rho(0)} =\left(\mathbb{1} + t \Q_{\S\I}  + \frac{t^2}{2!}\Q_{\S\I}^2   + \frac{t^3}{3!}\Q_{\S\I}^3 + \ldots  \right) \ket{\rho(0)}\,.
\end{equation}
The matrix $\Q_{\S\I}$ is a bilocal operator, which acts by matrix multiplication on the individual nodes in $\ket{\rho(0)}$. So the expansion can be seen as repeated applications of this operator on the initial state. Since $\Q_{\S\I}$ is constructed out of $P_\I$ and $\hSI$, we consider their action on the individual nodes first.  To do so, we introduce a graphical notation for the network of nodes. We will denote an infected node as a white circle
\begin{equation}
\begin{array}{c}
\begin{tikzpicture}
\node[shape=circle,draw=black] (A) at (0,0) {};
\end{tikzpicture}
\end{array} 
= \ket{\I} \,.
\end{equation}
The action of $P_\I$ and $\hSI$ on an infected node is then
\begin{equation} 
P_\I  \begin{array}{c}	
\begin{tikzpicture}
\begin{scope}
\node[shape=circle,draw=black] (A) at (0,0) {};
\end{scope}
\end{tikzpicture}
\end{array} =  \begin{array}{c}	
\begin{tikzpicture}
\begin{scope}
\node[shape=circle,draw=black] (A) at (0,0) {};
\end{scope}
\end{tikzpicture}
\end{array} \,, \qquad
\hSI \begin{array}{c}	
\begin{tikzpicture}
\begin{scope}
\node[shape=circle,draw=black] (A) at (0,0) {};
\end{scope}
\end{tikzpicture}
\end{array} = 0\,.
\end{equation}
We see that the infected state is an eigenvector for the infection projector with eigenvalue 1, meaning that projecting an infected to its infected component is a trivial operation. The infinitesimally stochastic matrix $\hSI$ acts as an annihilator on the infected state. This means that an infected node cannot be infected again.

When $P_\I$ and $\hSI$ act on any of the initially susceptible nodes, the following happens
\begin{equation}
P_\I \ket{\S} = 0 \,, \qquad \hSI \ket{\S} = \left( \begin{array}{c} -1 \\ 1 \end{array} \right) =  \ket{\I} - \ket{\S} \,.
\end{equation}
We see that $P_\I$ annihilates the susceptible state, meaning that susceptibles have zero infected probability. Acting with $\hSI$ on a susceptible creates an infinitesimally stochastic state whose components sum to zero. We will denote it graphically by a blue node:
\begin{equation}
\begin{array}{c}
\begin{tikzpicture}
\node[shape=circle,draw=black,fill=blue!20] (A) at (0,0) {};
\end{tikzpicture}
\end{array} 
= 
\ket{\I} -\ket{\S} 
\end{equation} 
The action of $P_\I$ and $\hSI$ on the infinitesimally stochastic state are
\begin{equation} 
P_\I  \begin{array}{c}	
\begin{tikzpicture}
\begin{scope}
\node[shape=circle,draw=black,fill=blue!20] (A) at (0,0) {};
\end{scope}
\end{tikzpicture}
\end{array} =  \begin{array}{c}	
\begin{tikzpicture}
\begin{scope}
\node[shape=circle,draw=black] (A) at (0,0) {};
\end{scope}
\end{tikzpicture}
\end{array} \,, \qquad
\hSI \begin{array}{c}	
\begin{tikzpicture}
\begin{scope}
\node[shape=circle,draw=black,fill=blue!20] (A) at (0,0) {};
\end{scope}
\end{tikzpicture}
\end{array} = - 
\begin{array}{c}	
\begin{tikzpicture}
\begin{scope}
\node[shape=circle,draw=black,fill=blue!20] (A) at (0,0) {};
\end{scope}
\end{tikzpicture}
\end{array} \,.
\end{equation}
This shows that blue nodes are eigenvectors of $\hSI$ with eigenvalue $-1$. $P_\I$ acts as a creation operator on the blue nodes; they turn blue nodes into white (infected) nodes, effectively creating a newly infected node. Hence, in order to create a newly infected from any of the initially susceptible states, they will first have to be acted upon with $\hSI$ to be turned into a infinitesimally stochastic state, followed by acting with $P_\I$ to turn the blue node white.

Finally, acting with $\Q_{\S\I}$ also introduces an edge from the adjacency matrix $A^{ij}$. In the graphical notation, these correspond to edges between the nodes. So we have at linear order in \eqref{timeexp} 
\begin{equation}\label{Hlinear}
\Q_{\S\I} \ket{\rho(0)} = \sum_{j \neq i}  { \scriptsize \begin{array}{c}
\begin{tikzpicture}
\node[shape=circle,draw=black] (A) at (0,0) {$i$};
\node[shape=circle,draw=black,fill=blue!20] (B) at (1,0) {$j$};

\path [-] (A) edge node[left] {} (B);
\end{tikzpicture}
\end{array}
}
\end{equation}
Here we have again set $\tau \to 1$ by scaling time $t \to t/\tau$. Node $i$ marks patient zero and after one application of $\Q_{\S\I}$ we see that it is in the process of infecting all its neighbors, which are now infinitesimally stochastic states. The remaining $N-2$ susceptible nodes in the network are not denoted explicitly here (and in the following), but we should always remember that they are present, waiting to be acted upon by $\Q_{\S\I}$.

In order to compute the probability of a node $k$ being infected, we have to compute the expectation value:
\begin{equation}
\vev{\I^k(t)} = \bra{1}P_\I^k \ket{\rho(t)}\,.
\end{equation}
This implies another operation of $P_\I^k$ on the system, followed by taking the $\ell^1$-norm. The property $\bra{1}\hSI = \bra{0}$ ensures that only diagrams without blue nodes will contribute to the final sum above. This implies that, apart from the initial conditions, the only contributions to the expectation value $\vev{\I^k}$ will come from when $P_\I^k$ acts on the blue node. At linear order, this gives explicitly:
\begin{equation}
\vev{\I^k(t)} = \delta^{ik} + t A^{ik}  + \cO(t^2)\,.
\end{equation}

The logistic differential equation \eqref{logisticeqn} shows exponential growth in the initial stage of an outbreak. This implies that the linear approximation will be valid only for a very short time. In order to make a better approximation, we have to consider computing the higher orders. From \eqref{timeexp} it is clear that the higher order terms are obtainable by repeated application of $\Q_{\S\I}$ on \eqref{Hlinear}. This process gives explicitly up to third order:
\begin{align}\label{rhoexp}
\ket{\rho(t)} = & \begin{array}{c}
\begin{tikzpicture}
\node[shape=circle,draw=black] (A) at (0,0) {};
\end{tikzpicture}
\end{array} + t 
\sum \begin{array}{c}
\begin{tikzpicture}
\node[shape=circle,draw=black] (A) at (0,0) {};
\node[shape=circle,draw=black,fill=blue!20] (B) at (.75,0) {};
\path [-] (A) edge node[left] {} (B);
\end{tikzpicture}
\end{array}
+ \frac{t^2}{2!} \sum 
\bigg[
\begin{array}{c}
\begin{tikzpicture}
\node[shape=circle,draw=black] (A) at (0,0) {};
\node[shape=circle,draw=black,fill=blue!20] (B) at (.75,-.5) {};
\node[shape=circle,draw=black,fill=blue!20] (C) at (.75,.5) {};
\path [-] (A) edge node[left] {} (B);
\path [-] (A) edge node[left] {} (C);
\end{tikzpicture}
\end{array}
+
\begin{array}{c}
\begin{tikzpicture}
\node[shape=circle,draw=black] (A) at (0,0) {};
\node[shape=circle,draw=black] (B) at (.75,0) {};
\node[shape=circle,draw=black,fill=blue!20] (C) at (1.5,0) {};
\path [-] (A) edge node[left] {} (B);
\path [-] (B) edge node[left] {} (C);
\end{tikzpicture}
\end{array}
-
\begin{array}{c}
\begin{tikzpicture}
\node[shape=circle,draw=black] (A) at (0,0) {};
\node[shape=circle,draw=black,fill=blue!20] (B) at (.75,0) {};
\path [-] (A) edge node[above] {\scriptsize 2} (B);
\end{tikzpicture}
\end{array} 
\bigg] \nonumber \\ &
+ \frac{t^3}{3!} \sum \bigg[ 
\begin{array}{c}
\begin{tikzpicture}
\node[shape=circle,draw=black] (A) at (0,0) {};
\node[shape=circle,draw=black,fill=blue!20] (B) at (.75,-.5) {};
\node[shape=circle,draw=black,fill=blue!20] (C) at (.75,.5) {};
\node[shape=circle,draw=black,fill=blue!20] (D) at (.75,0) {};
\path [-] (A) edge (B);
\path [-] (A) edge (C);
\path [-] (A) edge (D);
\end{tikzpicture}
\end{array}
+
3
\begin{array}{c}
\begin{tikzpicture}
\node[shape=circle,draw=black] (A) at (0,0) {};
\node[shape=circle,draw=black] (B) at (.75,0.5) {};
\node[shape=circle,draw=black,fill=blue!20] (C) at (.75,-.5) {};
\node[shape=circle,draw=black,fill=blue!20] (D) at (1.5,0.5) {};
\path [-] (A) edge node[left] {} (B);
\path [-] (A) edge node[left] {} (C);
\path [-] (B) edge (D);
\end{tikzpicture}
\end{array}
+
\begin{array}{c}
\begin{tikzpicture}
\node[shape=circle,draw=black] (A) at (0,0) {};
\node[shape=circle,draw=black] (B) at (.75,0) {};
\node[shape=circle,draw=black,fill=blue!20] (C) at (1.5,-0.5) {};
\node[shape=circle,draw=black,fill=blue!20] (D) at (1.5,0.5) {};
\path [-] (A) edge node[left] {} (B);
\path [-] (B) edge node[left] {} (C);
\path [-] (B) edge node[left] {} (D);
\end{tikzpicture}
\end{array}
+
\begin{array}{c}
\begin{tikzpicture}
\node[shape=circle,draw=black] (A) at (0,0) {};
\node[shape=circle,draw=black] (B) at (.75,0) {};
\node[shape=circle,draw=black] (C) at (1.5,0) {};
\node[shape=circle,draw=black,fill=blue!20] (D) at (2.25,0) {};
\path [-] (A) edge (B);
\path [-] (B) edge (C);
\path [-] (C) edge (D);
\end{tikzpicture}
\end{array} \nonumber \\ & 
- 3
\begin{array}{c}
\begin{tikzpicture}
\node[shape=circle,draw=black] (A) at (0,0) {};
\node[shape=circle,draw=black,fill=blue!20] (B) at (.75,-.5) {};
\node[shape=circle,draw=black,fill=blue!20] (C) at (.75,.5) {};
\path [-] (A) edge  (B);
\path [-] (A) edge node[above] {\scriptsize 2} (C);
\end{tikzpicture}
\end{array}
-
\begin{array}{c}
\begin{tikzpicture}
\node[shape=circle,draw=black] (A) at (0,0) {};
\node[shape=circle,draw=black] (B) at (.75,0) {};
\node[shape=circle,draw=black,fill=blue!20] (C) at (1.5,0) {};
\path [-] (A) edge node[above] {\scriptsize 2} (B);
\path [-] (B) edge node[left] {} (C);
\end{tikzpicture}
\end{array}
-
\begin{array}{c}
\begin{tikzpicture}
\node[shape=circle,draw=black] (A) at (0,0) {};
\node[shape=circle,draw=black] (B) at (.75,0) {};
\node[shape=circle,draw=black,fill=blue!20] (C) at (1.5,0) {};
\path [-] (A) edge node[left] {} (B);
\path [-] (B) edge node[above] {\scriptsize 2} (C);
\end{tikzpicture}
\end{array}
- 3
\begin{array}{c}
\begin{tikzpicture}
\node[shape=circle,draw=black] (A) at (0,0) {};
\node[shape=circle,draw=black] (B) at (.75,-.5) {};
\node[shape=circle,draw=black,fill=blue!20] (C) at (.75,.5) {};
\path [-] (A) edge (B);
\path [-] (A) edge (C);
\path [-] (B) edge (C);
\end{tikzpicture}
\end{array}
+ 
\begin{array}{c}
\begin{tikzpicture}
\node[shape=circle,draw=black] (A) at (0,0) {};
\node[shape=circle,draw=black,fill=blue!20] (B) at (.75,0) {};
\path [-] (A) edge node[above] {\scriptsize 3} (B);
\end{tikzpicture}
\end{array}
\bigg] \nonumber \\  & + \cO(t^4) \,.
\end{align}
Here in each term, the sum denotes the sum over all subgraph configurations of the given type, involving patient 0 as the left-most white node. Whenever a number appears above an edge, this means that the corresponding adjacency matrix entry $A^{ij}$ appears raised to the power of that number. If the network is unweighted, all non-zero entries to the adjacency matrix are unity, so the number can be ignored, but in principle \eqref{rhoexp} holds for weighted graphs as well. 

To compute the contribution to each order in $t$ from \eqref{rhoexp}, we need two pieces of information. We have to know how often a given subgraph appears in the network under consideration, which will depend on the underlying network structure. Another piece of information is universal to the $\Q_{\S\I}$ transition and it counts the coefficients of the different subgraphs. For instance, the loop term appears at third order with coefficient $-3$. This number is valid for any network and counts the number of different ways this subgraph can be constructed from 3 applications of $\Q_{\S\I}$ on the initial conditions. This property makes it possible to find recursive relations for these coefficients. To this end, consider how $\Q_{\S\I}$ would act on a diagram with an arbitrary number of white and blue nodes. There are four possible contributions, which constitute four rules universal to the $\StoI$ transition:
\begin{enumerate}
	\item When $\hSI$ acts on a new susceptible and $P_\I$ acts on a white node, a new blue node is added to the diagram and connected to the white node with an edge. For example:
	\begin{equation}\label{firstrule}
	\begin{array}{c}
	\begin{tikzpicture}
	\node[shape=circle,draw=black] (A) at (0,0) {};
	\node[shape=circle,draw=black,fill=blue!20] (B) at (.75,0) {};
	\path [-] (A) edge node[left] {} (B);
	\end{tikzpicture}
	\end{array} \to
	\begin{array}{c}
	\begin{tikzpicture}
	\node[shape=circle,draw=black] (A) at (0,0) {};
	\node[shape=circle,draw=black,fill=blue!20] (B) at (.75,-.5) {};
	\node[shape=circle,draw=black,fill=blue!20] (C) at (.75,.5) {};
	\path [-] (A) edge (B);
	\path [-] (A) edge (C);
	\end{tikzpicture}
	\end{array}
	\end{equation}
	\item When $\hSI$ acts on a new susceptible and $P_\I$ acts on a blue node, the blue node turns white and a new blue node is added to the diagram and connected to the (now white) existing node. For example:
	\begin{equation}\label{secondrule}
	\begin{array}{c}
	\begin{tikzpicture}
	\node[shape=circle,draw=black] (A) at (0,0) {};
	\node[shape=circle,draw=black,fill=blue!20] (B) at (.75,0) {};
	\path [-] (A) edge node[left] {} (B);
	\end{tikzpicture}
	\end{array}  \to 
	\begin{array}{c}
	\begin{tikzpicture}
	\node[shape=circle,draw=black] (A) at (0,0) {};
	\node[shape=circle,draw=black] (B) at (.75,0) {};
	\node[shape=circle,draw=black,fill=blue!20] (C) at (1.5,0) {};
	\path [-] (A) edge node[left] {} (B);
	\path [-] (B) edge node[left] {} (C);
	\end{tikzpicture}
	\end{array}
	\end{equation}
	
	\item When $\hSI$ acts on a blue nodes and $P_\I$ acts on a white node, a minus sign appears and an (additional) edge between the blue and white nodes is added. For example:
	\begin{equation}\label{thirdrule}
	\begin{array}{c}
	\begin{tikzpicture}
	\node[shape=circle,draw=black] (A) at (0,0) {};
	\node[shape=circle,draw=black,fill=blue!20] (B) at (.75,0) {};
	\path [-] (A) edge node[left] {} (B);
	\end{tikzpicture}
	\end{array}  \to - 
	\begin{array}{c}
	\begin{tikzpicture}
	\node[shape=circle,draw=black] (A) at (0,0) {};
	\node[shape=circle,draw=black,fill=blue!20] (B) at (.75,0) {};
	\path [-] (A) edge node[above] {\scriptsize 2} (B);
	\end{tikzpicture}
	\end{array} 
	\end{equation}
	\item When $\hSI$ acts on a blue node and $P_\I$ acts on another blue node, a minus sign appears and the second blue node is turned white. In addition, an edge between the two nodes is added. For example:
	\begin{equation}\label{fourthrule}
	\begin{array}{c}
	\begin{tikzpicture}
	\node[shape=circle,draw=black] (A) at (0,0) {};
	\node[shape=circle,draw=black,fill=blue!20] (B) at (.75,-.5) {};
	\node[shape=circle,draw=black,fill=blue!20] (C) at (.75,.5) {};
	\path [-] (A) edge node[left] {} (B);
	\path [-] (A) edge node[left] {} (C);
	\end{tikzpicture}
	\end{array} \to - 2
	\begin{array}{c}
	\begin{tikzpicture}
	\node[shape=circle,draw=black] (A) at (0,0) {};
	\node[shape=circle,draw=black] (B) at (.75,-.5) {};
	\node[shape=circle,draw=black,fill=blue!20] (C) at (.75,.5) {};
	\path [-] (A) edge (B);
	\path [-] (A) edge (C);
	\path [-] (B) edge (C);
	\end{tikzpicture}
	\end{array}
	\end{equation}
	Here the factor of 2 accounts for the two possible ways of applying $\Q$ to the two blue nodes.
\end{enumerate}
These four rules serve to interpret $\Q_{\S\I}$ as a subgraph-constructing algorithm. The first two rules govern adding new nodes and the last two rules specify adding internal edges to existing diagrams. If the number of possible subgraphs is constrained by choosing a regular or symmetric network structure, it becomes possible to solve the recursive relations imposed on the subgraphs and find exact solutions for the expectation value of infected. We give the details of this computation for regular trees and the circle graph in the next section.

Note that, while the expectation value $\vev{\I^k(t)}$ only receives contributions from diagrams with a single blue node, we are wrong to ignore multiple blue node diagrams in \eqref{rhoexp}. This is because the fourth rule above can remove a blue node and turn a diagram with two blue nodes into a single blue node diagram at the next order. This process necessarily involves loops in the network, since the blue nodes of any diagram are connected to patient 0 and the fourth rule connects two of them, forming a loop. In reverse, for tree graphs, the fourth rule can be safely ignored, as there are no loops, greatly simplifying the computation.\footnote{The third rule has the potential to generate loops as well, but it additionally raises the power on existing edges (with alternating sign), so it cannot be ignored completely for trees.} In that case, only tracking terms with a single blue node is sufficient for computing single node expectation values. If one is interested in correlation functions between two nodes, then terms with two blue nodes do need to be taken into account.

\section{Exact solutions for regular graphs}
\label{sec:exact}

In this section, we show explicitly how the procedure outlined in the previous section can be used to find exact solutions for networks with a regular structure. In particular, we will discuss the $\S\I$ outbreak from a single infected node on tree graphs and the cycle graph.

\subsection{Trees}
\label{sec:trees}

Lets construct a tree graph by following the spread from patient 0 outwards. Initially, there is only a single infected node. We define a distance variable $d$, which we take to be 0 for this initial configuration of a single node. Then we suppose that this infected node is capable of spreading the infection to $k_0$ other nodes. So at distance $d=1$, we have a star network of $k_0+1$ nodes, with the initially infected node in the center. Then we suppose that each of these $k_0$ nodes is capable of spreading the infection to $k_1 -1$ new nodes. At distance $d=2$, we thus have $1+ k_0 + k_0 (k_1-1)$ nodes in the network. For higher $d$, we continue to add $k_1-1$ new nodes to each node. The total number of nodes in a network of maximal distance $d_{\rm max}$ is
\begin{equation}\label{Nd}
N_{d_{\rm max}} = 1 + \sum_{n=1}^{d_{\rm max}} k_0 (k_1 -1)^{n-1} = 1 + \frac{k_0}{k_1-2} \left( (k_1 - 1)^{d_{\rm max}} - 1\right)
\end{equation}
An example of such a graph with $k_0 = 3 = k_1$ with  $d_{\rm max}=2$ is the Cayley tree:
\begin{equation}
\nonumber
\begin{tikzpicture}
\begin{scope}
\node[shape=circle,draw=black] (0) at (0,0) {};
\node[shape=circle,draw=black] (B) at (0,1) {};
\node[shape=circle,draw=black] (C) at (-.86,-.5) {};
\node[shape=circle,draw=black] (D) at (.86,-.5) {};
\node[shape=circle,draw=black] (E) at (-.86,1.5) {};
\node[shape=circle,draw=black] (F) at (.86,1.5) {};
\node[shape=circle,draw=black] (G) at (-1.73,0) {};
\node[shape=circle,draw=black] (H) at (-.86,-1.5) {};
\node[shape=circle,draw=black] (I) at (1.73,0) {};
\node[shape=circle,draw=black] (J) at (.86,-1.5) {};
\end{scope}

\begin{scope}
\path[-] (0) edge (B);
\path[-] (0) edge (C);
\path[-] (0) edge (D);
\path[-] (B) edge (E);
\path[-] (B) edge (F);
\path[-] (C) edge (G);
\path[-] (C) edge (H);
\path[-] (D) edge (I);
\path[-] (D) edge (J);
\end{scope}
\end{tikzpicture}
\end{equation}
The limit of infinite population size for $k_0 = k_1 = z$ gives a Bethe lattice of coordination number $z$ \cite{bethe1935statistical} . 

By constructing a graph in this way, we will never generate loops. This is advantageous, because to compute the single node expectation values we only have to track diagrams in the expansion \eqref{rhoexp} with exactly one blue node. In addition, the rules of the last section imply that the only terms appearing in the expansion \eqref{rhoexp} are chains with patient $0$ as the left-most node and a single blue node as the right-most node. If we restrict ourselves to unweighted graphs, this implies the expansion \eqref{rhoexp} on any tree can be organized as
\begin{equation}\label{rhoexptree}
\ket{\rho(t)} = \ket{\rho(0)} 
+ \sum_{n=0}^{\infty} \sum_{k=2}^{d_{\rm max}+1}  \frac{t^n}{n!}  \, a^k[n] 
{\scriptsize	
	\begin{tikzpicture}
	\node[shape=circle,draw=black] (A) at (0,0) {};
	\node[] (C) at (.4,0.1) {$\stackrel{k-1}{\ldots}$};
	\node[shape=circle,draw=black] (B) at (.8,0) {};
	\node[shape=circle,draw=black,fill=blue!20] (C) at (1.1,0) {};
	\end{tikzpicture}
} + \ldots
\end{equation}
Where ${\scriptsize	
	\begin{tikzpicture}
	\node[shape=circle,draw=black] (A) at (0,0) {};
	\node[] (C) at (.4,0.1) {$\stackrel{k-1}{\ldots}$};
	\node[shape=circle,draw=black] (B) at (.8,0) {};
	\node[shape=circle,draw=black,fill=blue!20] (C) at (1.1,0) {};
	\end{tikzpicture}
} $
denotes a chain of $k$ nodes with patient 0 on the left end and an infinitesimally stochastic state on the right end. The coefficients $a^k[n]$ simultaneously track the network dependent contributions (by counting the number of nodes at a distance $k-1$) and network independent contributions (by counting how many ways the chain subgraph can be obtained from acting $n$ times with $\Q_{\S\I}$ on the initial conditions). The parameter $d_{\rm max}$ denotes the maximal distance this chain can have in the case of finite trees. The dots denote terms with more than one blue node, which we will neglect for now. 

Acting with the transition matrix $\Q_{\S\I}$ on any given diagram gives the rules of section \ref{sec:perturbative}. From these rules, the first rule will generate a diagram with two blue nodes and the fourth rule only applies when there are loops in the network, so we will ignore these rules. The remaining two rules imply that
\begin{equation}\label{diagramrule}
\Q_{\rm \S\I} \cdot {\scriptsize	
	\begin{tikzpicture}
	\node[shape=circle,draw=black] (A) at (0,0) {};
	\node[] (C) at (.4,0.1) {$\stackrel{k-1}{\ldots}$};
	\node[shape=circle,draw=black] (B) at (.8,0) {};
	\node[shape=circle,draw=black,fill=blue!20] (C) at (1.1,0) {};
	\end{tikzpicture}
} =
-
{\scriptsize	
	\begin{tikzpicture}
	\node[shape=circle,draw=black] (A) at (0,0) {};
	\node[] (C) at (.4,0.1) {$\stackrel{k-1}{\ldots}$};
	\node[shape=circle,draw=black] (B) at (.8,0) {};
	\node[shape=circle,draw=black,fill=blue!20] (C) at (1.1,0) {};
	\end{tikzpicture}
}
+ (k_1 -1)
{\scriptsize	
	\begin{tikzpicture}
	\node[shape=circle,draw=black] (A) at (0,0) {};
	\node[] (C) at (.4,0.1) {$\stackrel{k-1}{\ldots}$};
	\node[shape=circle,draw=black] (B) at (.8,0) {};
	\node[shape=circle,draw=black] (C) at (1.1,0) {};
	\node[shape=circle,draw=black,fill=blue!20] (D) at (1.4,0) {};
	\end{tikzpicture}
}
\end{equation}
Here the factor $k_1-1$ comes from the graph specific property that each node with $d>0$ has degree $k_1$, so there are $k_1-1$ possible ways to attach a new blue node. To obtain a recursive relation for the coefficients $a^k[n]$ in \eqref{rhoexptree}, it is instructive to think about how a diagram at any given order can be obtained from the diagrams at a lower order. It is obvious from \eqref{diagramrule} that the contributions to $a^k[n]$ come from chains of length $k-1$ at order $n-1$ with multiplicity $(k_1 -1)$, and chains of length $k$ at order $n-1$ with a minus sign.
This leads to the recursive relation:
\begin{equation}\label{treerecrel}
a^k[n] = - a^k[n-1] + (k_1-1) a^{k-1}[n-1] \,,
\end{equation}
with the initial and boundary conditions 
\begin{equation}
a^2[1] = k_0 \,, \qquad a^k[n<k-1] = 0\,, \qquad  a^1[n] = 0 = a^{d_{\rm max}+2}[n]\,.
\end{equation}
The solution of this system is:
\begin{equation}\label{akntree}
a^k[n] =(-1)^{n-k+1} \frac{(n-1)!}{(k-2)! (n-k+1)!}  k_0 (k_1- 1)^{k-2} \,.
\end{equation}
These coefficients can be understood as follows. The alternating sign is present due to the third rule \eqref{thirdrule}. The factor $k_0(k_1-1)^{k-2}$ is the graph specific information, which counts the number of nodes at distance $d =k-1$ from patient 0. 
The combinatorial factor in \eqref{akntree} counts in how many ways a chain of length $k$ can be obtained from the initial conditions by $n$ repeated applications of $\Q_{\S\I}$. Each application of $\Q_{\S\I}$ can be thought of as adding an edge (with weight 1). So we should count how many ways there are to distribute $n$ edges between $k$ nodes on the condition that the resulting subgraph is a chain. The first $k-1$ edges will be used to form the chain. The remaining $n-k+1$ edges will have to be distributed among those $k-1$ edges. This is equivalent to distributing $n-k+1$ identical objects into $k-1$ distinct bins, for which there are $\left( \begin{array}{cc}
n-1 \\ k-2
\end{array}\right)$ ways of doing so.

To compute the expectation values for a single node at distance $d$ to be infected at a time $t$, we act with $P_\I$ on this node and compute the $\ell^1$-norm. We should also be careful to divide by the number of nodes at this distance, as \eqref{akntree} accounts for all nodes at a given distance. The answer is expressible as
\begin{equation}\label{Id}
\vev{\I^d(t)} = \frac{1}{k_0(k_1-1)^{d-1}} \sum_{n = d}^\infty a^{d+1}[n] \frac{t^n}{n!}  = 1 - \frac{\Gamma(d,t)}{\Gamma(d)} \,,
\end{equation}
where $\Gamma(d,t)$ is the upper incomplete Gamma function
\begin{equation}\label{Gammadef}
\Gamma(d,t) = \int_t^\infty s^{d-1} e^{-s} ds = (d-1)! e^{-t} \sum_{n=0}^{d-1} \frac{t^n}{n!} \,.
\end{equation}
To find the total number of infected as a function of time, we have to sum $\vev{\I^d(t)}$ times the number of nodes at distance $d$ over all $d$ and add one for the initially infected node:
\begin{equation}\label{Ittree1}
\vev{\I(t)}_{\rm tree} = 1 + \sum_{d=1}^{d_{\rm max}} k_0 (k_1 -1)^{d-1} \left(1- \frac{\Gamma(d,t)}{\Gamma(d)} \right)\,.
\end{equation}
Note that because the result \eqref{Id} only depends on the distance between the node of interest and patient 0, it is simple to find the total infected for a generic tree graph. One only has to replace the factor $k_0 (k_1 -1)^{d-1}$ in \eqref{Ittree1} with the number of nodes a distance $d$ from patient 0. 
The limit of large $N$ for the regular tree is obtained by taking $d_{\rm max} \to \infty$. In this limit, we can write
\begin{align}\label{IttreelargeN}
\lim_{N\to \infty} \vev{\I(t)}_{\rm tree} & = \sum_{a=1}^{\infty} \bra{1} P_\I^a  \ket{\rho(t)} = 1 +  \sum_{n=1}^{\infty} \sum_{k=2}^{n+1} \frac{t^n}{n!} a^k[n] \nonumber \\
& =   1 + \frac{k_0}{k_1-2} ( e^{(k_1-2)t} -1 )\,. 
\end{align}
The limit of $k_1 \to 2$ gives the special case where each of the $k_0$ branches attached to patient 0 are infinite line graphs. In that case the infected probability grows linearly as $1+ k_0 t$ , as can be verified by taking the $k_1 \to 2$ limit of the right hand side of \eqref{IttreelargeN}. When additionally $k_0 = 2$ this reduces to the infinite line graph.

\begin{figure}
	\centering
	\includegraphics[width=.6\textwidth]{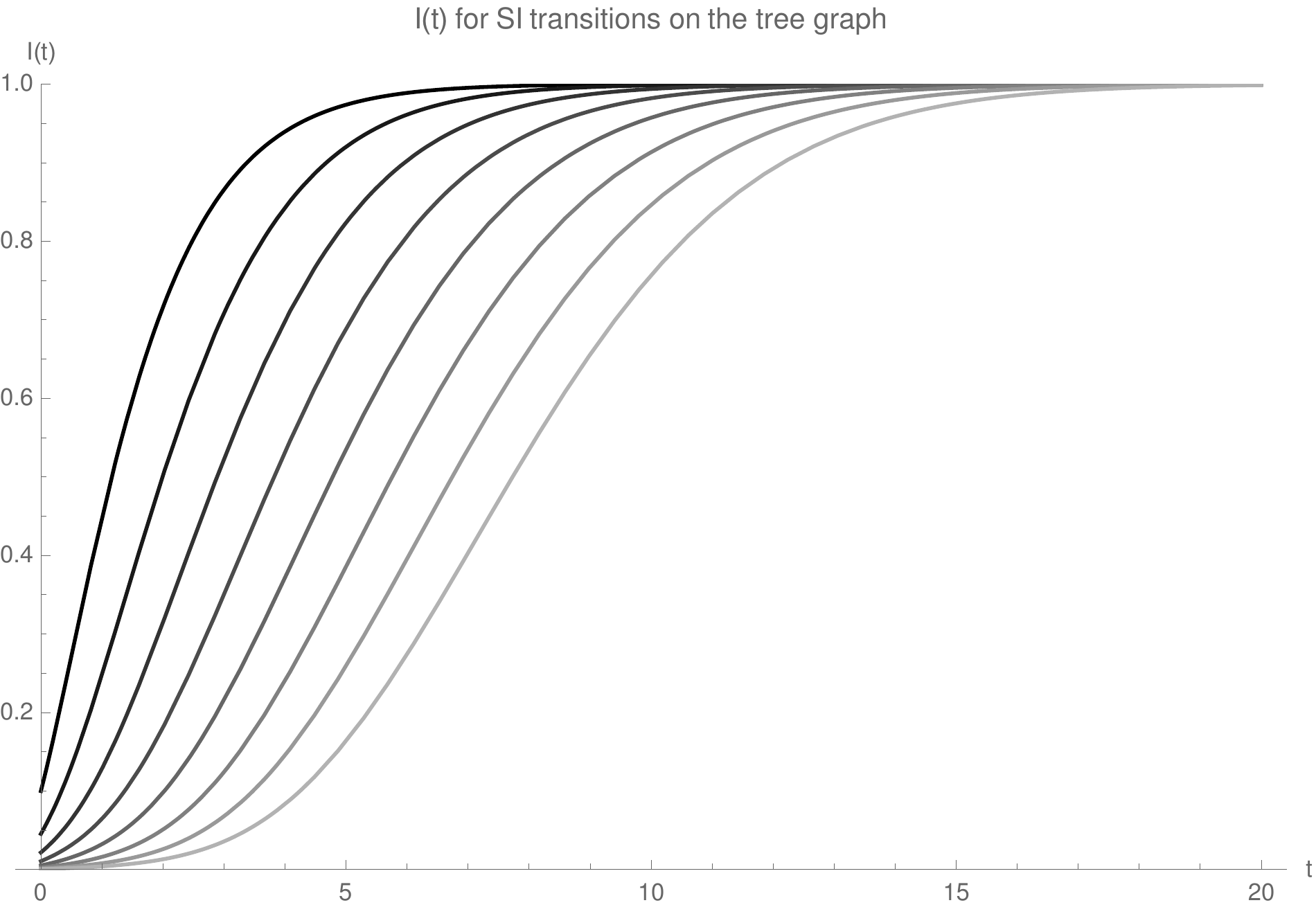}
	\caption{The total infected probability for the tree graph with $k_0 = 3 = k_1$ and $d_{\rm max}= 2,3,4,5,6,7,8,9$ (such that $N_{d_{\rm max}} = 10, 22,46,94,190,382, 766, 1534 $) from darkest to lightest shading of gray.}
	\label{fig:tree}
\end{figure}

For any higher $k_1$ value, the infected probability grows exponentially. This is, of course, due to the fact that each newly infected node is able to infect more than one other node, leading to exponential growth. The exponential growth is only damped when the full system size is reached. Starting at the order $n=d_{\rm max}+1$, there are no new diagrams to add at any higher order in the expansion. The fact that each individual chain converges to a finite value as \eqref{Id}, leads to the convergence of the total number of infected in the finite graph case. Some exact solutions for the total infected probability for finite $N_{d_{\rm max}}$ and $k_0 = 3 = k_1$ have been plotted in figure \ref{fig:tree}.

\subsection{Ring}
\label{sec:rings}

To illustrate the additional complication of loops in the network, we will consider here the simplest network containing a loop: the circle graph, or  ring. Because the ring is translation invariant, it does not matter where on the ring we place patient 0. The infection can spread in two directions, clockwise and counter-clockwise. Any newly infected node can only infect one other node, such that the expected infection prevalence grows linearly. The limit of large $N$ should reproduce the results of the infinite line graph of the last subsection:
\begin{equation}\label{ringlargeN}
\lim_{N\to \infty} \vev{\I(t)}_{\rm ring} = 1+ 2t \,.
\end{equation}
For finite $N$, we have to track contributions from the loop diagram and from diagrams with two blue nodes, as they will eventually meet up to form the loop. This implies that now we can write the expansion \eqref{rhoexp} as
\begin{align}\label{rhoexpring}
\ket{\rho(t)} = \ket{\rho(0)} + \sum_{n=1}^\infty & \frac{t^n}{n!} \left( \sum_{k=2}^{N} \bigg( a^k[n] 
{\scriptsize	
	\begin{tikzpicture}
	\node[shape=circle,draw=black] (A) at (0,0) {};
	\node[] (C) at (.4,0.1) {$\stackrel{k-1}{\ldots}$};
	\node[shape=circle,draw=black] (B) at (.8,0) {};
	\node[shape=circle,draw=black,fill=blue!20] (C) at (1.1,0) {};
	\end{tikzpicture}
}
+ b^k[n]
{\scriptsize	
	\begin{tikzpicture}
	\node[shape=circle,draw=black,fill=blue!20] (D) at (0,0) {};		
	\node[shape=circle,draw=black] (A) at (0.3,0) {};
	\node[] (C) at (.7,0.1) {$\stackrel{k-2}{\ldots}$};
	\node[shape=circle,draw=black] (B) at (1.1,0) {};
	\node[shape=circle,draw=black,fill=blue!20] (C) at (1.4,0) {};
	\end{tikzpicture}
}
\right)  \\
& \nonumber \qquad \qquad  + a^{\rm loop}[n]
{\scriptsize	
	\begin{tikzpicture}
	\node[shape=circle,draw=black] (A) at (0,0) {};
	\node[] (C) at (.4,0.1) {$\stackrel{N-1}{\ldots}$};
	\node[shape=circle,draw=black] (B) at (.8,0) {};
	\node[shape=circle,draw=black,fill=blue!20] (C) at (1.1,0) {};
	\path[-] (A) edge[bend left=90]  (C);
	\end{tikzpicture}
}
\bigg)\,.
\end{align}
Following the rules of section \ref{sec:perturbative}, we see that the following transitions are imposed by acting with $\Q_{\S\I}$ on the diagrams in \eqref{rhoexpring}
\begin{subequations}
\begin{align}
\Q_{\S\I} \cdot {\scriptsize	
	\begin{tikzpicture}
	\node[shape=circle,draw=black] (A) at (0,0) {};
	\node[] (C) at (.4,0.1) {$\stackrel{k-1}{\ldots}$};
	\node[shape=circle,draw=black] (B) at (.8,0) {};
	\node[shape=circle,draw=black,fill=blue!20] (C) at (1.1,0) {};
	\end{tikzpicture}
} = & 
- {\scriptsize	
	\begin{tikzpicture}
	\node[shape=circle,draw=black] (A) at (0,0) {};
	\node[] (C) at (.4,0.1) {$\stackrel{k-1}{\ldots}$};
	\node[shape=circle,draw=black] (B) at (.8,0) {};
	\node[shape=circle,draw=black,fill=blue!20] (C) at (1.1,0) {};
	\end{tikzpicture}
}
+
{\scriptsize	
	\begin{tikzpicture}
	\node[shape=circle,draw=black] (A) at (0,0) {};
	\node[] (C) at (.4,0.1) {$\stackrel{k-1}{\ldots}$};
	\node[shape=circle,draw=black] (B) at (.8,0) {};
	\node[shape=circle,draw=black] (C) at (1.1,0) {};
	\node[shape=circle,draw=black,fill=blue!20] (D) at (1.4,0) {};
	\end{tikzpicture}
}
+
{\scriptsize	
	\begin{tikzpicture}
	\node[shape=circle,draw=black,fill=blue!20] (D) at (0,0) {};		
	\node[shape=circle,draw=black] (A) at (0.3,0) {};
	\node[] (C) at (.7,0.1) {$\stackrel{k-1}{\ldots}$};
	\node[shape=circle,draw=black] (B) at (1.1,0) {};
	\node[shape=circle,draw=black,fill=blue!20] (C) at (1.4,0) {};
	\end{tikzpicture}
}\,, & \forall k < N \\
\Q_{\S\I} \cdot {\scriptsize	
	\begin{tikzpicture}
	\node[shape=circle,draw=black,fill=blue!20] (D) at (0,0) {};		
	\node[shape=circle,draw=black] (A) at (0.3,0) {};
	\node[] (C) at (.7,0.1) {$\stackrel{k-2}{\ldots}$};
	\node[shape=circle,draw=black] (B) at (1.1,0) {};
	\node[shape=circle,draw=black,fill=blue!20] (C) at (1.4,0) {};
	\end{tikzpicture}
} = &
- 2 \,
{\scriptsize	
	\begin{tikzpicture}
	\node[shape=circle,draw=black,fill=blue!20] (D) at (0,0) {};		
	\node[shape=circle,draw=black] (A) at (0.3,0) {};
	\node[] (C) at (.7,0.1) {$\stackrel{k-2}{\ldots}$};
	\node[shape=circle,draw=black] (B) at (1.1,0) {};
	\node[shape=circle,draw=black,fill=blue!20] (C) at (1.4,0) {};
	\end{tikzpicture}
}
+ 2 \, 
{\scriptsize	
	\begin{tikzpicture}
	\node[shape=circle,draw=black,fill=blue!20] (D) at (0,0) {};		
	\node[shape=circle,draw=black] (A) at (0.3,0) {};
	\node[] (C) at (.7,0.1) {$\stackrel{k-1}{\ldots}$};
	\node[shape=circle,draw=black] (B) at (1.1,0) {};
	\node[shape=circle,draw=black,fill=blue!20] (C) at (1.4,0) {};
	\end{tikzpicture}
}\,, & \forall k < N \\
\Q_{\S\I} \cdot {\scriptsize	
	\begin{tikzpicture}
	\node[shape=circle,draw=black] (A) at (0,0) {};
	\node[] (C) at (.4,0.1) {$\stackrel{N-1}{\ldots}$};
	\node[shape=circle,draw=black] (B) at (.8,0) {};
	\node[shape=circle,draw=black,fill=blue!20] (C) at (1.1,0) {};
	\end{tikzpicture}
} = & 
- {\scriptsize	
	\begin{tikzpicture}
	\node[shape=circle,draw=black] (A) at (0,0) {};
	\node[] (C) at (.4,0.1) {$\stackrel{N-1}{\ldots}$};
	\node[shape=circle,draw=black] (B) at (.8,0) {};
	\node[shape=circle,draw=black,fill=blue!20] (C) at (1.1,0) {};
	\end{tikzpicture}
}
-
{\scriptsize	
	\begin{tikzpicture}
	\node[shape=circle,draw=black] (A) at (0,0) {};
	\node[] (C) at (.4,0.1) {$\stackrel{N-1}{\ldots}$};
	\node[shape=circle,draw=black] (B) at (.8,0) {};
	\node[shape=circle,draw=black,fill=blue!20] (C) at (1.1,0) {};
	\path[-] (A) edge[bend left=90]  (C);
	\end{tikzpicture}
} \,,
\\
\Q_{\S\I} \cdot {\scriptsize	
	\begin{tikzpicture}
	\node[shape=circle,draw=black,fill=blue!20] (D) at (0,0) {};		
	\node[shape=circle,draw=black] (A) at (0.3,0) {};
	\node[] (C) at (.7,0.1) {$\stackrel{N-2}{\ldots}$};
	\node[shape=circle,draw=black] (B) at (1.1,0) {};
	\node[shape=circle,draw=black,fill=blue!20] (C) at (1.4,0) {};
	\end{tikzpicture}
} = &
- 2 \,
{\scriptsize	
	\begin{tikzpicture}
	\node[shape=circle,draw=black,fill=blue!20] (D) at (0,0) {};		
	\node[shape=circle,draw=black] (A) at (0.3,0) {};
	\node[] (C) at (.7,0.1) {$\stackrel{N-2}{\ldots}$};
	\node[shape=circle,draw=black] (B) at (1.1,0) {};
	\node[shape=circle,draw=black,fill=blue!20] (C) at (1.4,0) {};
	\end{tikzpicture}
}
- 2 \, 
{\scriptsize	
	\begin{tikzpicture}
	\node[shape=circle,draw=black] (A) at (0,0) {};
	\node[] (C) at (.4,0.1) {$\stackrel{N-1}{\ldots}$};
	\node[shape=circle,draw=black] (B) at (.8,0) {};
	\node[shape=circle,draw=black,fill=blue!20] (C) at (1.1,0) {};
	\path[-] (A) edge[bend left=90]  (C);
	\end{tikzpicture}
}\,,  \\
\Q_{\S\I} \cdot {\scriptsize	
	\begin{tikzpicture}
	\node[shape=circle,draw=black] (A) at (0,0) {};
	\node[] (C) at (.4,0.1) {$\stackrel{N-1}{\ldots}$};
	\node[shape=circle,draw=black] (B) at (.8,0) {};
	\node[shape=circle,draw=black,fill=blue!20] (C) at (1.1,0) {};
	\path[-] (A) edge[bend left=90]  (C);
	\end{tikzpicture}
}
= & 
-2 \,
{\scriptsize	
	\begin{tikzpicture}
	\node[shape=circle,draw=black] (A) at (0,0) {};
	\node[] (C) at (.4,0.1) {$\stackrel{N-1}{\ldots}$};
	\node[shape=circle,draw=black] (B) at (.8,0) {};
	\node[shape=circle,draw=black,fill=blue!20] (C) at (1.1,0) {};
	\path[-] (A) edge[bend left=90]  (C);
	\end{tikzpicture}
} \,,
\end{align}
\end{subequations}
We see that the diagrams with two blue nodes are only able to become diagrams with one blue node once the whole ring has been infected. 
Once again, we can obtain recursive relations for the coefficients $a^k[n], b^k[n]$ and $a^{\rm loop}[n]$. This can be done by listing the coefficients of subgraphs at order $n-1$, which transition into the sought-after diagrams after another application of $\Q_{\S\I}$. The result is:
\begin{subequations}\label{ringrecrel}
	\begin{align}
	a^k[n] & = - a^k[n-1] + a^{k-1}[n-1] \,, \\
	b^k[n] & = -2 b^k[n-1] +2 b^{k-1}[n-1] + a^{k-1}[n-1] \,, \\
	a^{\rm loop}[n] & = - 2 a^{\rm loop} [n-1] - 2 b^N[n-1] - a^N[n-1] \,. 
	\end{align}
\end{subequations}
The initial and boundary conditions for these recursion relations are:
\begin{equation}
a^{k}[n<k-2] = 0 \,, \qquad b^k[n<k-2] = 0\,, \qquad b^{k<3}[n] = 0
\end{equation}
as well as
\begin{equation}
a^2[1] = 2\,,
\end{equation}
reflecting that patient 0 can infect two other nodes.
We see that the first recursion relation in \eqref{ringrecrel} only depends on the coefficients $a^k[n]$. Its solution is \eqref{akntree} with $k_0=2=k_1$, or:
\begin{equation}\label{asol}
a^k[n] = (-1)^{n+k-1} \frac{2 (n-1)!}{(k-2)! (n+1-k)!} \,.
\end{equation}
Using this in the second recursion relation enables us to find $b^k[n]$ as
\begin{equation}\label{bsol}
b^k[n] = (-1)^{n+k-1} \frac{(2^n - 2) (n-2)!}{(k-3)!(n-k+1)!} \,.
\end{equation}
Finally, using \eqref{asol} and \eqref{bsol} in the recursion relation for $a^{\rm loop}[n]$ and solving it gives
\begin{equation}\label{loopsol}
a^{\rm loop}[n] = (-1)^{n+N-1}  \frac{(2^n -2)(n-2)!}{(N-2)!(n-N)!} \,.
\end{equation}
Notice that $a^{\rm loop}[n] = - b^{N+1}[n]$, which implies that the contribution from $N$-node loop diagrams is equal to minus the contribution of $N+1$ node chains with blue nodes at both ends. The latter term does not contribute to single node expectation values, but it does give a contribution to the correlation function of two nodes.

\begin{figure}
	\centering
	\includegraphics[width=.6\textwidth]{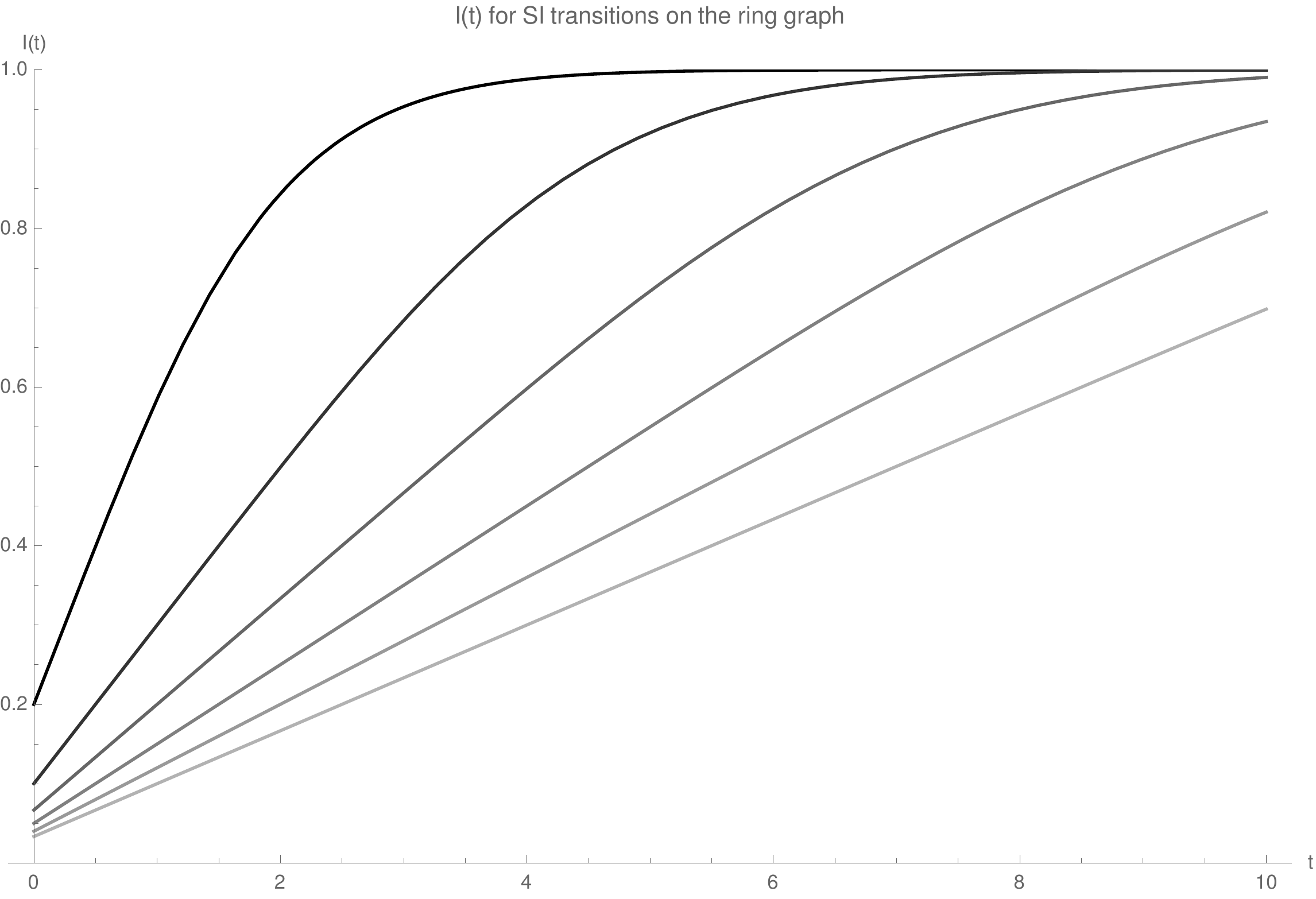}
	\caption{The total infected probability for the ring graph with $N= 5, 10,15,20,25,30$ from darkest to lightest shading of gray.}
	\label{fig:ring}
\end{figure}

We can now obtain the expectation value for the total number of infected for the $N$ node ring graph by summing all coefficients corresponding to terms with a single blue node. This sum can be organized as
\begin{equation}\label{Itring1}
\vev{\I(t)}_{\rm ring} = 1 + \sum_{n=1}^{N-1} \frac{t^n}{n!} \sum_{k=2}^{n+1} a^k[n] + \sum_{n=N}^\infty \frac{t^n}{n!} \left( \sum_{k=2}^N a^k[n] + a^{\rm loop}[n]\right) \,.
\end{equation}
The first sum in this expression only receives a contribution from $n=1$. To see this, we express it as a sum over the binomial coefficient with alternating sign.
\begin{equation}\label{akfixed}
\sum_{k=2}^{n+1} a^k[n] = 2 \sum_{k=1}^n (-1)^{n+k}  \left(\begin{array}{c}
n-1 \\
k-1
\end{array}\right) = 2 \delta_{n,1}
\end{equation}
Here $\delta_{n,1} = 1$ if $n=1$ and 0 otherwise. This sum is over all entries of a given row of Pascals triangle, but with alternating sign. Each row will cancel, except for the line with a single entry at the top of the triangle, which is 1. This allows us to verify immediately the $N \to \infty$ limit of \eqref{Itring1}. In that limit, the first sum in \eqref{Itring1} is the only contributing term and \eqref{akfixed} ensures that only the linear term contributes with coefficient 2, in agreement with \eqref{ringlargeN}. 
But the solution \eqref{Itring1} holds for any $N$ and we have found explicit expressions for the coefficients $a^k[n]$ and $a^{\rm loop}[n]$. The result \eqref{Itring1} can be expressed in terms of the incomplete Gamma function \eqref{Gammadef} as
\begin{equation}
\label{Itringexpl}
\frac{1}{N} \vev{\I(t)}_{\rm ring} = 1 - \frac{(2t)^N e^{-2t} }{N!} +  \frac{(1-N+2t) }{N!} \Gamma(N,2t)\,.
\end{equation}
We have plotted the total infected probability for some values of $N$ in figure \eqref{fig:ring}. It clearly shows linear growth until the full capacity of the system is reached, when the system stabilizes to the completely infected state.

The two examples treated in this section show how viewing the exact transition rate matrix as a sum of bilocal operators on the tensor product space of individual probability vectors can be used to find exact solutions for logistic growth on simple graphs. It is interesting to ask whether this can be extended to other networks, like the complete graph or to random networks with specified degree distribution and local clustering coefficients. In addition, for epidemiological models including recovery, the rules of section \ref{sec:perturbative} will have to be extended to allow for the recovery process. This means that the number of contributing diagrams will increase considerably. In that case, it might no longer be possible to solve the recursive relations as neatly as in the examples treated here. It could then be advantageous to use probability generating functions to generate the coefficients at a given order. The use of generating functions is well-established in the network science community and gives the advantage to incorporate network properties such as the degree distribution and clustering. We leave an exploration along these lines to future work.

\section{Conclusion and Discussion}
\label{sec:conclusion}

In this paper we have developed a tensor product formulation for obtaining exact transition rate matrices for the continuous-time Markovian dynamics of the epidemic spreading process on networks. We have seen how the infinitesimal generator for the exact Markov chain can be understood as a sum over bilocal operators acting on individual probability vector spaces representing each node in the network. The state of the population is an element of the tensor product space of the individual probability vector spaces. The more familiar node-based dynamics on networks is recovered by restricting to separable (direct product) states and the non-linear dynamics of this mean-field approximation is a consequence of the bilocality of the exact transition rate matrices. The construction gives insight into the microscopic dynamics of the epidemic spreading process and is suitable for adapting mathematical techniques used in quantum many-body systems to the study of information spread on complex networks.

As an example of how the exact transition rate matrices give insight in the microscopic dynamics of the spreading process, we have analyzed the time dependence of an $\S\I$ outbreak on graphs. We found that the transition rate matrix can be interpreted as a subgraph generating algorithm for the perturbative expansion in time. This allows one to derive recursive relations for the coefficients contributing to each order in the expansion and we found exact solutions of epidemic outbreaks on tree graphs and the cycle graph. It is conceivable that this approach can be extended to find exact solutions for $\S\I$ outbreaks on other types of regular graphs or for random networks constructed from a configuration model.

For more sophisticated models of epidemiology including recovery, using this approach to obtain exact solutions of outbreaks on finite graphs will be more difficult, as the rules imposed by the transition rate matrices are more complicated. In these cases, it might be advantageous to adapt numerical and analytical methods from the study of many-body quantum spin systems or quantum information to stochastic systems on networks. 
In recent years, significant progress has been made in simulating strongly correlated quantum spin systems by tensor networks \cite{verstraete2008matrix,orus2014practical}. The use of tensor networks has been explored in a variety of different disciplines, such as quantum chemistry \cite{szalay2015tensor},  machine learning \cite{glasser2018supervised}, as well as for the description of many-body non-equilibrium stochastic systems \cite{blythe2007nonequilibrium,Temme2010,Johnson2010,Hotta2015,Harada2019}. They are efficient ways of representing composite states as networks of matrices, contracted in specific ways. The size of these matrices represents the amount of correlation between the constituents of the system. The formalism developed in this paper is well suited for application to stochastic tensor networks, both for studying the steady state properties and for explicit time-evolution. It would be interesting to investigate how tensor networks can be used to further our understanding of stochastic systems on complex networks.

\subsection{Measurements and memory}
\label{sec:measurements}

The approach followed in this paper gives the exact infinitesimal generators of the many-body Markov chain, so in principle $S(t) = \exp(\Q t)$ contains information of all past correlations in the system. It tracks all ways the infection can spread through the network. Still, the defining feature of a Markovian system is that the transition rates into any state only depend on the state of the system at that time. We show here how these two statements are compatible and how measurements on subsystems can serve as memory for an otherwise memory-less Markovian system. 

An equivalent and frequently used definition of a Markovian system is the statement that conditional probabilities do not depend on past interactions
\begin{equation}\label{Markovdef}
P \big( \Y_n, t_n \big| \Y_{n-1}, t_{n-1} ; \ldots ;  \Y_{1}, t_1 \big)  =   P \big( \Y_{n-1}, t_{n-1} \big| \Y_{n-1} ,t_{n-1} \big) \,.
\end{equation}
The conditional probability $ P( \Y_{n-1}, t_{n-1} | \Y_{n-1} ,t_{n-1})$ of finding the system in the microscopic state $\Y_n \in C^N$ at time $t_n$, given that it was in the state $\Y_{n-1} $ at time $t_{n-1}$ is computed as:
\begin{equation}\label{condprob}
P \big(  \Y_n, t_n \big| \Y_{n-1} t_{n-1}  \big) =
\frac{\bra{1} P_{\Y_n} S(t_n-t_{n-1}) P_{\Y_{n-1}} \ket{\rho(t_{n-1})}}{\bra{1} P_{\Y_{n-1}} \ket{\rho(t_{n-1})} }
\end{equation}
Here $P_{\Y_n} = \ket{\Y_n}\bra{\Y_n} $ is the projection operator which projects the system to the microscopic configuration $\Y_n$ and $S(t) = \exp(\Q t)$. 
If we suppose that $\ket{\rho(t)}$ has non-zero probability of being in the state $\Y_{n-1}$ at time $t_{n-1}$, then
\begin{equation}
P_{\Y_{n-1}} \ket{\rho(t_{n-1})} = \bracket{\Y_{n-1}}{\rho(t_{n-1})} \, \ket{\Y_{n-1}}\,.
\end{equation}
The conditioning on this state is performed by normalizing the projection
\begin{equation}
\frac{P_{\Y_{n-1}} \ket{\rho(t_{n-1})} }{\bracket{\Y_{n-1}}{\rho(t_{n-1})} } = \ket{\Y_{n-1}}\,.
\end{equation}
The normalized projection hence `collapses' the system into a particular, deterministic microstate, much like in quantum measurements and the associated wave function collapse. It is this property that ensures that \eqref{Markovdef} is satisfied, as long as all projections leading up to $t_{n-1}$ are non-zero. The conditional probabilities \eqref{condprob} are then easily found to be the matrix elements of $S(t_n- t_{n-1})$:
\begin{equation}
 P ( \Y_{n-1}, t_{n-1} | \Y_{n-1} ,t_{n-1}) = \bra{\Y_n} S(t_n- t_{n-1}) \ket{\Y_{n-1}} \,.
\end{equation}

The microscopic description is also capable of computing conditional probabilities between any two individual nodes, or subsets of nodes. These are computed not by projecting the state $\ket{\rho(t)}$ to a specific deterministic state $\Y_n$ of the full network, but by projecting only on the nodes of interest. 

Say we divide the population into two groups, $A $ and $B$. At a given time $t_1$, we perform a measurement on the subset $A$. This can be implemented by projecting $\ket{\rho(t_1)}$ on the measurement outcome for nodes in the subset $A$, but acting with identity operator on the nodes in the complement $B$. Conditioning on this projected state produces a new stochastic state:
\begin{equation}\label{projection}
\frac{P_{\Y_A} \otimes \mathbb{1}_B \ket{\rho(t_1)}}{\bra{1}P_{\Y_A} \otimes \mathbb{1}_B \ket{\rho(t_1)}  } = \ket{\Y_A} \otimes \ket{ \rho_B(t_1)} \,.
\end{equation}
Where now $\Y_A \in C^{N_A}$ and $N_A$ is the number of nodes in the set $A$. Now, the projection is only partial and produces a state depending explicitly on $t_1$, the time at which this operation is performed. This means the property \eqref{Markovdef} does not hold for taking partial projections and the system retains a memory of the past interactions.

The partial projections can be used in a practical setting. If one has managed to construct a representation of $\ket{\rho(t)}$ (for instance by using a tensor network) to model the state of a population at a given time, then measurements on real world individuals can be used to improve the model by projecting as \eqref{projection}. The state then contains a more accurate description of the measured subset, while retaining the previous correlations within the subset $B$. Furthermore, the expectation values of individuals in the subset $B$ will adjust automatically due to the measurement outcome on $A$, increasing the accuracy of the model. This could proof useful in modeling systems where only partial information on the state of the system is available, as is often the case with epidemic outbreaks.

\subsection{Final thoughts}
\label{sec:finalthoughts}

The transition rate matrices constructed in this paper describe the continuous time Markovian dynamics of the spreading process on static networks. Hence the applicability is limited by the assumption of the spreading process being Markovian and the network topology being static. In recent years, generalizations to non-Markovian spreading processes and time-dependent (adaptive) networks have become popular, see for instance \cite{karsai2012universal,van2013non,karsai2014time,PastorSatorras2015,jo2014analytically}. However, even in the case of non-Markovian dynamics, the Markovian description is still useful, as it has been shown in \cite{starnini2017equivalence} that the steady state properties of the $\SIS$ model with non-Markovian transmission can be well approximated by the Markovian model with an effective transmission rate, regardless of the underlying network topology.

Generalizations of the work in this paper to time-varying networks and non-Poissonian interevent distributions are possible. One could imagine that instead of evolving the system with $\exp(\Q t)$ for static networks, where individuals are continuously connected, one could decompose the time evolution operator into a product of smaller exponential operators, akin to the Trotter-Suzuki decomposition \cite{suzuki1990fractal,nielsen2002quantum}. The order and duration of applying these smaller exponential operators could then vary as a function of time, with transition rates chosen from heavy-tailed probability distributions. The state of the system $\ket{\rho(t)}$ can then be constructed by patching together nodes with logistic gates defined in section \ref{sec:Logisticgate}. This approach is similar to performing a computation in a quantum circuit model. 

Many possible generalizations of the current approach to other stochastic systems with many interacting constituents come to mind. Several binary Markovian systems on complex networks have been described in \cite{gleeson2013binary} and these are well-suited for a description in the framework presented here. We could also think of generalizations to multiple species and bipartite graphs to define dynamical ecological systems (such as preditor-prey models) on complex networks exactly. It is conceivable that this work can be useful for voter-models \cite{sood2005voter}, rumor spreading \cite{daley1965stochastic}, and many other dynamical models which have become popular in the study of complex systems \cite{boccaletti2006complex,castellano2009statistical}. Finally, models of neural activity in the brain \cite{ginzburg1994theory,george2009towards,bullmore2009complex} follow similar stochastic dynamical rules as those explored in this paper and we hope that developing mathematical tools along the lines sketched here might eventually contribute to the understanding of neuronal dynamics.

\begin{acknowledgements}
The author thanks Jay Armas, Yael Artzy-Randrup, Philippe Corboz, Shyam Gopalakrishnan, Ido Niessen, Piet Van Mieghem, and Michael Walter for useful discussions and comments. The author is especially grateful to Ivano Lodato for countless inspiring conversations and extensive critical comments. The author also gratefully acknowledges the hospitality provided by the Physique Th\'eorique et Math\'ematique and the International Solvay Institutes during the completion of this work.
\end{acknowledgements}

%
\section*{Conflict of interest}

The authors declare that they have no conflict of interest.

\appendix
\section{Some properties of tensor products}
\label{sec:appendixA}

This appendix collects some properties of tensor products for readers unfamiliar with the subject. As mentioned in the main text, the tensor product $\otimes:  V, W \mapsto V \otimes W$ is a bilinear map from two vector spaces $V, W$ to the product space $V \otimes W$, which is a $\text{Dim}(V) \times \text{Dim}(W)$-dimensional vector space. It is a generalization of the outer product to tensors of generic rank. In terms of tensor products of matrices in a specified basis, one can think of it as the Kronecker product of the two matrices. For instance, in the case of 2-dimensional matrices
\begin{align}
A \otimes B & = \left( \begin{array}{cc}
a_{11} & a_{12} \\
a_{21} & a_{22}
\end{array}\right) \otimes
 \left( \begin{array}{cc}
b_{11} & b_{12} \\
b_{21} & b_{22}
\end{array}\right)  = 
\left(\begin{array}{cc}
a_{11}B & a_{12}B \\
a_{21}B & a_{22}B
\end{array}\right) 
\\ \nonumber
& = 
\left(\begin{array}{cccc}
a_{11}b_{11} & a_{11}b_{12} & a_{12} b_{11} & a_{12} b_{12} \\
a_{11}b_{21} & a_{11}b_{22} & a_{12} b_{21} & a_{12} b_{22} \\
a_{21}b_{11} & a_{21}b_{12} & a_{22} b_{11} & a_{22} b_{12} \\
a_{21}b_{21} & a_{21}b_{22} & a_{22} b_{21} & a_{22} b_{22} 
\end{array}\right) 
\end{align}
By definition, the tensor product is bilinear, i.e. linear in both arguments
\begin{align}
(A + B) \otimes C & = A \otimes C + B \otimes C \,, \\
 C \otimes (A + B) & = C \otimes A + C \otimes B\,, \\
c (A \otimes B) & = (c A) \otimes B = A \otimes ( c B) \,,
\end{align}
where $A,B,C$ are arbitrary tensors and $c$ is a scalar.

The usefulness of the tensor product description stems from the fact that it allows one to formulate the dynamics in terms of local operations on the constituent vector spaces. This follows from the property that tensor products also operate on linear maps between vector spaces. Specifically, if $A$ and $B$ are linear maps $A: V \to X$ and $B: W \to Y$, then the tensor product $A \otimes B$ is a linear map $A\otimes B : V\otimes W \to X \otimes Y$. Practically, this implies for matrix multiplications of tensor products that:
\begin{equation}\label{localop}
(A \otimes B)  (C \otimes D) = (A  C) \otimes (B  D)\,,
\end{equation}
where $A, B,C,D$ are arbitrary tensors, constraint such that the pairs $A,C$ and $B, D$ are contractible. A direct consequence of this property is the vanishing of the $\ell^1$-norm of any tensor product involving at least one infinitesimally stochastic matrix (defined as having zero column sums). Indeed if $\bra{1} \h =0$, then also 
\begin{equation}
\bra{1} (A \otimes \h) = \bra{1} A \otimes \bra{1} \h = 0\,.
\end{equation}


\begin{thebibliography}{10}
	\providecommand{\url}[1]{{#1}}
	\providecommand{\urlprefix}{URL }
	\expandafter\ifx\csname urlstyle\endcsname\relax
	\providecommand{\doi}[1]{DOI~\discretionary{}{}{}#1}\else
	\providecommand{\doi}{DOI~\discretionary{}{}{}\begingroup
		\urlstyle{rm}\Url}\fi
	
	\bibitem{albert2002statistical}
	Albert, R., Barabási, A.L.: Statistical mechanics of complex networks.
	\newblock Reviews of modern physics \textbf{74}(1), 47 (2002)
	
	\bibitem{Anderson1992}
	Anderson, R.M., May, R.M.: Infectious diseases of humans.
	\newblock Oxford University Press, Oxford (1992)
	
	\bibitem{ben1992mean}
	ben Avraham, D., Köhler, J.: Mean-field (n, m)-cluster approximation for
	lattice models.
	\newblock Physical Review A \textbf{45}(12), 8358 (1992)
	
	\bibitem{Baez_2017}
	Baez, J.C., Biamonte, J.: Quantum techniques for stochastic mechanics.
	\newblock arXiv preprint arXiv:1209.3632  (2012)
	
	\bibitem{bailey1950simple}
	Bailey, N.T.: A simple stochastic epidemic.
	\newblock Biometrika p. 193–202 (1950)
	
	\bibitem{bancal2010steady}
	Bancal, J.D., Pastor-Satorras, R.: Steady-state dynamics of the forest fire
	model on complex networks.
	\newblock The European Physical Journal B \textbf{76}(1), 109–121 (2010)
	
	\bibitem{barrat2008dynamical}
	Barrat, A., Barthelemy, M., Vespignani, A.: Dynamical processes on complex
	networks.
	\newblock Cambridge university press (2008)
	
	\bibitem{bethe1935statistical}
	Bethe, H.A.: Statistical theory of superlattices.
	\newblock Proceedings of the Royal Society of London. Series A-Mathematical and
	Physical Sciences \textbf{150}(871), 552–575 (1935)
	
	\bibitem{blythe2007nonequilibrium}
	Blythe, R.A., Evans, M.R.: Nonequilibrium steady states of matrix-product form:
	a solver's guide.
	\newblock Journal of Physics A: Mathematical and Theoretical \textbf{40}(46),
	R333 (2007)
	
	\bibitem{boccaletti2006complex}
	Boccaletti, S., Latora, V., Moreno, Y., Chavez, M., Hwang, D.U.: Complex
	networks: Structure and dynamics.
	\newblock Physics reports \textbf{424}(4-5), 175–308 (2006)
	
	\bibitem{Brauer2012}
	Brauer, F., Castillo-Chavez, C., Castillo-Chavez, C.: Mathematical models in
	population biology and epidemiology, vol.~2.
	\newblock Springer (2012)
	
	\bibitem{bullmore2009complex}
	Bullmore, E., Sporns, O.: Complex brain networks: graph theoretical analysis of
	structural and functional systems.
	\newblock Nature reviews neuroscience \textbf{10}(3), 186–198 (2009)
	
	\bibitem{castellano2009statistical}
	Castellano, C., Fortunato, S., Loreto, V.: Statistical physics of social
	dynamics.
	\newblock Reviews of modern physics \textbf{81}(2), 591 (2009)
	
	\bibitem{cator2013susceptible}
	Cator, E., {Van Mieghem}, P.: Susceptible-infected-susceptible epidemics on the
	complete graph and the star graph: Exact analysis.
	\newblock Physical Review E \textbf{87}(1), 012811 (2013)
	
	\bibitem{daley1965stochastic}
	Daley, D.J., Kendall, D.G.: Stochastic rumours.
	\newblock IMA Journal of Applied Mathematics \textbf{1}(1), 42–55 (1965)
	
	\bibitem{derrida1993exact}
	Derrida, B., Evans, M.R., Hakim, V., Pasquier, V.: Exact solution of a 1d
	asymmetric exclusion model using a matrix formulation.
	\newblock Journal of Physics A: Mathematical and General \textbf{26}(7), 1493
	(1993)
	
	\bibitem{felderhof1971spin}
	Felderhof, B.: Spin relaxation of the ising chain.
	\newblock Reports on Mathematical Physics \textbf{1}(3), 215–234 (1971)
	
	\bibitem{ganesh2005effect}
	Ganesh, A., Massoulié, L., Towsley, D.: The effect of network topology on the
	spread of epidemics.
	\newblock In: Proceedings IEEE 24th Annual Joint Conference of the IEEE
	Computer and Communications Societies., vol.~2, p. 1455–1466. IEEE (2005)
	
	\bibitem{george2009towards}
	George, D., Hawkins, J.: Towards a mathematical theory of cortical
	micro-circuits.
	\newblock PLoS Comput Biol \textbf{5}(10), e1000532 (2009)
	
	\bibitem{ginzburg1994theory}
	Ginzburg, I., Sompolinsky, H.: Theory of correlations in stochastic neural
	networks.
	\newblock Physical review E \textbf{50}(4), 3171 (1994)
	
	\bibitem{glasser2018supervised}
	Glasser, I., Pancotti, N., Cirac, J.I.: Supervised learning with generalized
	tensor networks.
	\newblock arXiv preprint arXiv:1806.05964  (2018)
	
	\bibitem{gleeson2013binary}
	Gleeson, J.P.: Binary-state dynamics on complex networks: Pair approximation
	and beyond.
	\newblock Physical Review X \textbf{3}(2), 021004 (2013)
	
	\bibitem{gleeson2012accuracy}
	Gleeson, J.P., Melnik, S., Ward, J.A., Porter, M.A., Mucha, P.J.: Accuracy of
	mean-field theory for dynamics on real-world networks.
	\newblock Physical Review E \textbf{85}(2), 026106 (2012)
	
	\bibitem{grassberger1983critical}
	Grassberger, P.: On the critical behavior of the general epidemic process and
	dynamical percolation.
	\newblock Mathematical Biosciences \textbf{63}(2), 157–172 (1983)
	
	\bibitem{Harada2019}
	Harada, K., Kawashima, N.: Entropy governed by the absorbing state of directed
	percolation.
	\newblock Physical review letters \textbf{123}(9), 090601 (2019)
	
	\bibitem{hethcote1981nonlinear}
	Hethcote, H.W., Stech, H.W., {Van Den Driessche}, P.: Nonlinear oscillations in
	epidemic models.
	\newblock SIAM Journal on Applied Mathematics \textbf{40}(1), 1–9 (1981)
	
	\bibitem{Hotta2015}
	Hotta, Y.: Tensor-network algorithm for nonequilibrium relaxation in the
	thermodynamic limit.
	\newblock Physical Review E \textbf{93}(6), 062136 (2016)
	
	\bibitem{jo2014analytically}
	Jo, H.H., Perotti, J.I., Kaski, K., Kertész, J.: Analytically solvable model
	of spreading dynamics with non-poissonian processes.
	\newblock Physical Review X \textbf{4}(1), 011041 (2014)
	
	\bibitem{Johnson2010}
	Johnson, T., Clark, S., Jaksch, D.: Dynamical simulations of classical
	stochastic systems using matrix product states.
	\newblock Physical Review E \textbf{82}(3), 036702 (2010)
	
	\bibitem{karsai2012universal}
	Karsai, M., Kaski, K., Barabási, A.L., Kertész, J.: Universal features of
	correlated bursty behaviour.
	\newblock Scientific reports \textbf{2}(1), 1–7 (2012)
	
	\bibitem{karsai2014time}
	Karsai, M., Perra, N., Vespignani, A.: Time varying networks and the weakness
	of strong ties.
	\newblock Scientific reports \textbf{4}, 4001 (2014)
	
	\bibitem{keeling1999effects}
	Keeling, M.J.: The effects of local spatial structure on epidemiological
	invasions.
	\newblock Proceedings of the Royal Society of London. Series B: Biological
	Sciences \textbf{266}(1421), 859–867 (1999)
	
	\bibitem{keeling2011modeling}
	Keeling, M.J., Rohani, P.: Modeling infectious diseases in humans and animals.
	\newblock Princeton university press (2011)
	
	\bibitem{kermack1927}
	Kermack, W.O., McKendrick, A.G.: A contribution to the mathematical theory of
	epidemics.
	\newblock Proceedings of the Royal Society \textbf{A}(115), 700–721 (1927)
	
	\bibitem{kermack1932contributions}
	Kermack, W.O., McKendrick, A.G.: Contributions to the mathematical theory of
	epidemics. ii.—the problem of endemicity.
	\newblock Proceedings of the Royal Society of London. Series A, containing
	papers of a mathematical and physical character \textbf{138}(834), 55–83
	(1932)
	
	\bibitem{kiss2017mathematics}
	Kiss, I.Z., Miller, J.C., Simon, P.L., et~al.: Mathematics of epidemics on
	networks, vol. 598.
	\newblock Springer (2017)
	
	\bibitem{kiss2015exact}
	Kiss, I.Z., Morris, C.G., Sélley, F., Simon, P.L., Wilkinson, R.R.: Exact
	deterministic representation of markovian sir epidemics on networks with and
	without loops.
	\newblock Journal of mathematical biology \textbf{70}(3), 437–464 (2015)
	
	\bibitem{kuperman2001small}
	Kuperman, M., Abramson, G.: Small world effect in an epidemiological model.
	\newblock Physical Review Letters \textbf{86}(13), 2909 (2001)
	
	\bibitem{liu2004spread}
	Liu, J., Tang, Y., Yang, Z.: The spread of disease with birth and death on
	networks.
	\newblock Journal of Statistical Mechanics: Theory and Experiment
	\textbf{2004}(08), P08008 (2004)
	
	\bibitem{ludwig1975final}
	Ludwig, D.: Final size distribution for epidemics.
	\newblock Mathematical Biosciences \textbf{23}(1-2), 33–46 (1975)
	
	\bibitem{masuda2010heterogeneous}
	Masuda, N., Gibert, N., Redner, S.: Heterogeneous voter models.
	\newblock Physical Review E \textbf{82}(1), 010103 (2010)
	
	\bibitem{may2001infection}
	May, R.M., Lloyd, A.L.: Infection dynamics on scale-free networks.
	\newblock Physical Review E \textbf{64}(6), 066112 (2001)
	
	\bibitem{nekovee2007theory}
	Nekovee, M., Moreno, Y., Bianconi, G., Marsili, M.: Theory of rumour spreading
	in complex social networks.
	\newblock Physica A: Statistical Mechanics and its Applications
	\textbf{374}(1), 457–470 (2007)
	
	\bibitem{newman2010networks}
	Newman, M.: Networks: An introduction.
	\newblock Oxford university press (2010)
	
	\bibitem{newman2003properties}
	Newman, M.E.: Properties of highly clustered networks.
	\newblock Physical Review E \textbf{68}(2), 026121 (2003)
	
	\bibitem{Newman2002}
	Newman, M.E.J.: Spread of epidemic disease on networks.
	\newblock Physical review \textbf{E}(66), 016128 (2002)
	
	\bibitem{nielsen2002quantum}
	Nielsen, M.A., Chuang, I.: Quantum computation and quantum information.
	\newblock Cambridge University Press (2010)
	
	\bibitem{orus2014practical}
	Orús, R.: A practical introduction to tensor networks: Matrix product states
	and projected entangled pair states.
	\newblock Annals of Physics \textbf{349}, 117–158 (2014)
	
	\bibitem{PastorSatorras2015}
	Pastor-Satorras, R., Castellano, C., {Van Mieghem}, P., Vespignani, A.:
	Epidemic processes in complex networks.
	\newblock Reviews of modern physics \textbf{87}(3), 925 (2015)
	
	\bibitem{pastor2001epidemic}
	Pastor-Satorras, R., Vespignani, A.: Epidemic spreading in scale-free networks.
	\newblock Physical review letters \textbf{86}(14), 3200 (2001)
	
	\bibitem{pastor2002epidemic}
	Pastor-Satorras, R., Vespignani, A.: Epidemic dynamics in finite size
	scale-free networks.
	\newblock Physical Review E \textbf{65}(3), 035108 (2002)
	
	\bibitem{poulin2011quantum}
	Poulin, D., Qarry, A., Somma, R., Verstraete, F.: Quantum simulation of
	time-dependent hamiltonians and the convenient illusion of hilbert space.
	\newblock Physical review letters \textbf{106}(17), 170501 (2011)
	
	\bibitem{rand1999correlation}
	Rand, D.: Correlation equations and pair approximations for spatial ecologies.
	\newblock Advanced ecological theory: principles and applications p. 100–142
	(1999)
	
	\bibitem{sahneh2013generalized}
	Sahneh, F.D., Scoglio, C., {Van Mieghem}, P.: Generalized epidemic mean-field
	model for spreading processes over multilayer complex networks.
	\newblock IEEE/ACM Transactions on Networking \textbf{21}(5), 1609–1620
	(2013)
	
	\bibitem{serrano2006percolation}
	Ángeles Serrano, M., Boguná, M.: Percolation and epidemic thresholds in
	clustered networks.
	\newblock Physical review letters \textbf{97}(8), 088701 (2006)
	
	\bibitem{simon2011exact}
	Simon, P.L., Taylor, M., Kiss, I.Z.: Exact epidemic models on graphs using
	graph-automorphism driven lumping.
	\newblock Journal of mathematical biology \textbf{62}(4), 479–508 (2011)
	
	\bibitem{sood2005voter}
	Sood, V., Redner, S.: Voter model on heterogeneous graphs.
	\newblock Physical review letters \textbf{94}(17), 178701 (2005)
	
	\bibitem{starnini2017equivalence}
	Starnini, M., Gleeson, J.P., Boguñá, M.: Equivalence between non-markovian
	and markovian dynamics in epidemic spreading processes.
	\newblock Physical review letters \textbf{118}(12), 128301 (2017)
	
	\bibitem{suzuki1990fractal}
	Suzuki, M.: Fractal decomposition of exponential operators with applications to
	many-body theories and monte carlo simulations.
	\newblock Physics Letters A \textbf{146}(6), 319–323 (1990)
	
	\bibitem{szalay2015tensor}
	Szalay, S., Pfeffer, M., Murg, V., Barcza, G., Verstraete, F., Schneider, R.,
	Örs Legeza: Tensor product methods and entanglement optimization for ab
	initio quantum chemistry.
	\newblock International Journal of Quantum Chemistry \textbf{115}(19),
	1342–1391 (2015)
	
	\bibitem{Temme2010}
	Temme, K., Verstraete, F.: Stochastic matrix product states.
	\newblock Phys. Rev. Lett. \textbf{104}, 210502 (2010)
	
	\bibitem{van2014exact}
	{Van Mieghem}, P.: Exact markovian sir and sis epidemics on networks and an
	upper bound for the epidemic threshold.
	\newblock arXiv preprint arXiv:1402.1731  (2014)
	
	\bibitem{van2008virus}
	{Van Mieghem}, P., Omic, J., Kooij, R.: Virus spread in networks.
	\newblock IEEE/ACM Transactions On Networking \textbf{17}(1), 1–14 (2008)
	
	\bibitem{van2013non}
	{Van Mieghem}, P., {Van de Bovenkamp}, R.: Non-markovian infection spread
	dramatically alters the susceptible-infected-susceptible epidemic threshold
	in networks.
	\newblock Physical review letters \textbf{110}(10), 108701 (2013)
	
	\bibitem{verhulst1838}
	Verhulst, P.F.: Notice sur la loi que la population suit dans son
	accroissement.
	\newblock Correspondance mathématique et physique \textbf{10}, 113–121
	(1838)
	
	\bibitem{verstraete2008matrix}
	Verstraete, F., Murg, V., Cirac, J.I.: Matrix product states, projected
	entangled pair states, and variational renormalization group methods for
	quantum spin systems.
	\newblock Advances in Physics \textbf{57}(2), 143–224 (2008)
	
	\bibitem{wang2003epidemic}
	Wang, Y., Chakrabarti, D., Wang, C., Faloutsos, C.: Epidemic spreading in real
	networks: An eigenvalue viewpoint.
	\newblock In: 22nd International Symposium on Reliable Distributed Systems,
	2003. Proceedings., p. 25–34. IEEE (2003)
	
	\bibitem{WEISS1971261}
	Weiss, G.H., Dishon, M.: On the asymptotic behavior of the stochastic and
	deterministic models of an epidemic.
	\newblock Mathematical Biosciences \textbf{11}(3), 261–265 (1971)
	
	\bibitem{zanette2002dynamics}
	Zanette, D.H.: Dynamics of rumor propagation on small-world networks.
	\newblock Physical review E \textbf{65}(4), 041908 (2002)
	
\end{thebibliography}

\end{document}